\documentclass[10pt]{article} % [11pt]
\usepackage{graphicx} % Required for inserting images
\usepackage[dvipsnames,svgnames,table]{xcolor}

\RequirePackage[T1]{fontenc} \RequirePackage[tt=false, type1=true]{libertine} 
\usepackage[a4paper,top=25mm,bottom=25mm,left=25mm,right=25mm]{geometry}
\usepackage{amsmath,amsthm,amssymb}
\usepackage{setspace} % To set line spacing
\usepackage{adjustbox}
\usepackage{longtable}

\usepackage{tabularx}
\newcolumntype{R}{>{\raggedleft\arraybackslash}X}
\newcolumntype{L}{>{\raggedright\arraybackslash}X}
\newcolumntype{C}{>{\centering\arraybackslash}X}
\usepackage[usestackEOL]{stackengine}
\usepackage{booktabs}
\usepackage{threeparttable}
\usepackage{multirow}
\usepackage[nottoc,notlot,notlof]{tocbibind} % includes everything in ToC

\usepackage[justification=centerfirst]{caption}
\usepackage[labelformat=simple]{subcaption}
\usepackage{alphalph}

\DeclareCaptionLabelFormat{parenthesis}{(#2)}
\makeatletter
\renewcommand\p@subfigure{\arabic{figure}.}
\makeatother

\usepackage{tikz}
\usetikzlibrary{arrows, calc, matrix, patterns, positioning, shapes, trees}

\newcommand\Pair[3]{%
  \begin{tabular}{|>{\centering\arraybackslash}m{0.75cm}|>{\centering\arraybackslash}m{1.5cm}|}
  \hline
  \multirow{2}{*}{#1} & #2 \\ \cline{2-2}
   & #3 \\
  \hline
  \end{tabular}%
}

\usepackage[hyphens]{xurl}

\usepackage{natbib}
\usepackage{appendix}

\usepackage{float}

\usepackage{hyperref}
\hypersetup{
    colorlinks=true,
    citecolor=Blue,
    linkcolor=Blue,
    filecolor=Plum,      
    urlcolor=CadetBlue,
    pdftitle={Overleaf Example},
    }

\usepackage{arydshln}

\makeatletter
\newcommand*{\rom}[1]{\expandafter\@slowromancap\romannumeral #1@}
\makeatother

\usepackage[normalem]{ulem}
\def\keywords{\vspace{.5em} 
{\noindent \textit{Keywords}: }}

\usepackage{authblk}
\title{Tournament design: A review from \\ an operational research perspective}
\author[1]{Karel Devriesere\footnote{Corresponding author, email: \texttt{karel.devriesere@ugent.be}}}
\author[2,3]{L\'aszl\'o Csat\'o}
\author[1,4]{Dries Goossens}
\affil[1]{~Faculty of Economics and Business Administration, Ghent University, Tweekerkenstraat
2, 9000 Ghent, Belgium}
\affil[2]{~Institute for Computer Science and Control (SZTAKI), Hungarian Research Network (HUN-REN), Hungary}
\affil[3]{~Corvinus University of Budapest, Hungary}
\affil[4]{~Core lab CVAMO, FlandersMake@UGent, Belgium}
\date{}   %% don't need date to appear
\begin{document}

\maketitle

\begin{abstract}
\noindent
Every sport needs rules. Tournament design refers to the rules that determine how a tournament, a series of games between a number of competitors, is organized. This study aims to provide an overview of the tournament design literature from the perspective of operational research. Three important design criteria are discussed: efficacy, fairness, and attractiveness. Our survey classifies the papers discussing these properties according to the main components of tournament design: format, seeding, draw, scheduling, and ranking. We also outline several open questions and promising directions for future research.
\end{abstract}

\keywords{OR in sports; ranking; scheduling; seeding; tournament design}

%\AMS{90-10, 90B35, 90B90, 91B14}
% Mathematical modeling or simulation for problems pertaining to operations research and mathematical programming
% Deterministic scheduling theory in operations research
% Case-oriented studies in operations research
% Social choice

%\JEL{C44, C63, Z20}
% Operations Research, Statistical Decision Theory
% Computational Techniques, Simulation Modeling 
% Sport Economics: General

%\listoftables

%\listoffigures

%\tableofcontents

%\revision{Added text}
%\sout{Deleted text}

\section{Introduction} \label{Sec1}

Practicing sports is an important part of the life of many people all over the world as many health, psychological, and social benefits are associated with physical activity. Watching and following sports is even more popular than playing it. About 3 billion people watched the 2020 Summer Olympic Games \citep{Statista2023b}, while no less than 1.5 billion people tuned in for the 2022 FIFA World Cup final \citep{FIFA2023WorldCup}. %These are just a few examples that show the involvement of fans all over the world with sports. In addition, many sports fans their identities are deeply entwined with the team or club they support (\cite{heere2007sports}).
Even tournament design aspects can be popular: the draw for the 2021/22 UEFA Champions League Round of 16 has attracted more than 1 million views on the YouTube channel of UEFA \citep{BoczonWilson2023}.
Hence, ``\emph{designing an optimal contest is both a matter of significant financial concern for the organizers, participating individuals, and teams, and a matter of consuming personal interest for millions of fans}'' \citep[p.~1137]{szymanski2003economic}.
%The Olympic games and FIFA World Cup are among the most popular sports tournaments in the world. In a sports tournament, individual or teams of athletes compete against each other. Next to providing entertainment for spectators and marketing purposes, one of the primary goals of a sports tournament is to determine the ``best'' player out of a set of contenders \citep{placek2023impossibility}.

Every tournament is governed by a set of rules that determine the set of contestants, the format, the schedule, the ranking, and the prize allocation. Deciding on these rules is what we call \emph{tournament design}. In the following, the contestants will be called players or teams interchangeably.
The set of matches to be played is given by the tournament \emph{format}. \emph{Seeding} is typically used to guarantee that a strong contestant is not eliminated at the early stages of the tournament merely because it plays against an even stronger opponent. Subject to these and other constraints, the contestants can be allocated randomly, through a so-called \emph{draw}. At first sight, the \emph{schedule}, the order and timing of the matches, seems a purely operational issue (involving travel distance, referee and venue availability, etc.). However, it may have powerful effects on the fairness and the outcome of the tournament.
\emph{Ranking} aims to order the teams according to their performance in the tournament. Although performance can be directly measured in some sports (e.g.\ by finishing times), most sports allow only to rank the contestants by paired comparisons, games or matches. \emph{Prize allocation} determines how prizes (e.g.\ promotion, relegation, qualification, broadcasting revenue) will be awarded to the contestants. 

Given the variety of tournament designs, the question arises on what basis they can be compared. We identify four different criteria that appear frequently in the literature on tournament design: 1) Efficacy; 2) Fairness; 3) Attractiveness; and 4) Strategy-proofness.
\emph{Efficacy} is defined as the ability of a tournament to reveal the ``true'' (but hidden) ranking of participants. In contrast, \emph{fairness} has no uniform definition (see Section~\ref{Sec232} and \citet{Baumer2024}); in our paper, it mainly refers to the principle of equal treatment. % Still, we organize the literature according to two common interpretations of fairness: (a) a player cannot be disadvantaged compared to any weaker player; or (b) all players should be treated equally.
\emph{Attractiveness} also has multiple interpretations, but they are always connected to the excitement that the tournament generates. Finally, all participants should have an incentive to exert full effort and play according to their true strength under any possible scenario in the tournament. That is, the tournament design needs to be \emph{strategy-proof}.

Our focus on these criteria may suggest that sports federations can freely choose an optimal tournament design from their own perspective, i.e., a design showing an appropriate balance of theoretical properties. In practice, however, traditions and the nature of the sport may play an even larger role in this decision process. For instance, European football leagues are usually played in a round-robin format (Section~\ref{Sec211}), while tennis competitions are organized as a seeded knockout tournament (Section~\ref{Sec212}). Similarly, the Swiss system (Section~\ref{Sec213}) is mainly used in chess. Sports differ in how much time a player needs to recover between two matches: for instance, baseball teams play almost daily, while World Rugby recommends that at least four clear recovery days are provided after each match. The number of rest days required may depend on the frequency of changes allowed by the rules, too; ice hockey teams can play every second day mainly because they often change lines.
Some sports plan their season in order avoid overlap with other popular sports \citep{Pagels2018}, while travel costs may force tournaments to be organized in a short time span on one location \citep{Nurmi2014}. Hence, any ``good'' tournament design depends on the particular sport, and operational research should take these constraints into account to improve decision making. However, in this paper -- apart from a number of specific examples -- we will maintain a more generic view.

Tournament design is a research field at the interface between operational research and (sports) economics. Operational research mainly contributes to tournament design by optimizing design choices over certain criteria and aiding decision-making. In sports economics, the main issue is typically how the tournament design may affect the behavior and efforts of the players, through theoretical or empirical research \citep{Medcalfe2024, Palacios-Huerta2023}. There is a clear synergy between these two angles: sports economics research can bring new criteria and conditions to be taken into account in optimization (or debunk myths that traditionally played a role in tournament design), and validate the decisions proposed by operational research.
% has received growing interest in the scientific literature over the last decades as reflected by our list of references. It

In this paper, we assume an operational research perspective. 
Therefore, prize allocation will be discussed only marginally, as this is, in our view, related to sports economics rather than to operational research. We refer the reader to \citet{MoldovanuSela2001}, \citet{bergantinos2020sharing}, and \cite{DietzenbacherKondratev2023} for excellent discussions on this topic. Furthermore, even though this is certainly one of the most important theoretical properties of tournament design, we have decided \emph{not} to review strategy-proofness here for the same reason. In addition, including this issue would have substantially increased the length of our survey, and a recent book deals with the problem of incentives from an operational research perspective \citep{Csato2021a}. Analogously, the effect of tournament design on competitive balance, consumer demand, or market structure will be mostly ignored; we refer the interested reader to a seminal economics paper on these aspects \citep{szymanski2003economic}. Nevertheless, we always try to emphasize practical applications by connecting the topics to real-world tournaments. As such, this survey will hopefully be useful for sports economists, too, and advance collaboration between the two domains. Sports administrators and tournament organizers can also benefit from such an up-to-date review of the tournament design literature.

Previous surveys on operational research and tournament design have so far mainly focused on scheduling, and, in particular, its operational aspects, see \citet{Rasmussen2008}, \citet{KendallKnustRibeiroUrrutia2010}, \citet{Ribeiro2012}, \citet{van2020robinx}, and \citet{RibeiroUrrutiadeWerra2023b}. Self-contained textbooks about sports scheduling are \citet{briskorn2008sports} and \citet{RibeiroUrrutiadeWerra2023a}. There are also good -- albeit somewhat outdated -- overviews of the operational research analysis of sports rules \citep{wright200950, wright2014or}. \citet{kendall2017sports} consider a number of examples from sports where rule changes have had unintended consequences.
The recent survey of \citet{lenten2022scholarly} collects academic works that proposed rule changes in sports.
We add to the existing surveys by considering a broader set of papers and covering a wider set of design issues and criteria, motivated by recent contributions of operational research in these areas. On the other hand, we offer less by excluding many studies that do not explicitly deal with tournament design but focus on, for example, individual games.
%This comprehensive -- but by no means exhaustive -- study attempts to overview many papers that analyse tournament design with the tools of operational research. Hopefully, both the academic community and sports administrators can benefit from such an up-to-date review of tournament design literature.

The rest of the paper is organized as follows. Section~\ref{Sec2} provides the necessary background and terminology on tournament formats, ranking, and the three design criteria presented above. Investigations of tournament formats, seeding policies, draw rules, scheduling, and ranking methods with respect to these criteria are discussed in Sections~\ref{Sec3}--\ref{Sec7}.
Finally, Section~\ref{Sec8} concludes and lists several open questions identified by us. 

\section{Basic terminology and background} \label{Sec2}

Tournament design uses several, not always well-defined, notions. This section attempts to summarize and clarify them in order to prepare and structure the following discussion. In particular, Section~\ref{Sec21} presents tournament formats, Section~\ref{Sec22} overviews ranking, and Section~\ref{Sec23} defines three criteria that play a crucial role in the evaluation of tournament designs.

\subsection{Tournament formats} \label{Sec21}

The format of a tournament indicates the set of matches to be played, without identifying the particular teams. Three fundamental types of tournament formats exist: knockout, round robin, and Swiss system tournaments. They are presented in Sections~\ref{Sec211}, \ref{Sec212}, and \ref{Sec213}, respectively. More complex tournament formats can be formed by combining these formats, as briefly discussed in Section~\ref{Sec214}.

\subsubsection{Round robin} \label{Sec211}

One of the basic competition formats is the \emph{round robin tournament} \citep{harary1966theory}. In a $k$-round robin ($k$RR) tournament, each participant faces each other $k$ times. Most domestic football leagues are $k$RR tournaments \citep{lasek2018efficacy}; for example, the competitions in the top five leagues (England, France, Germany, Italy, Spain) are organized as a 2RR. Since 1RR is typically called ``single round robin'' and 2RR is called ``double round robin'', we will use the notations SRR and DRR, respectively, for these formats.

If not all rounds of a round robin tournament are played, by design (especially if there are many contestants), or because the tournament is aborted (see Section~\ref{Sec721}), we call it an \emph{incomplete round robin tournament}. While this format dates back to at least \citet{cochran1971designs}, who discussed it in the context of bridge, it got more attention through the qualifying phase of the 2019/20 CONCACAF Nations League, which was contested by 34 teams in a single group, playing just 4 matches each (two home and two away). UEFA adopted an incomplete round robin structure for its club competitions from the 2024/25 season: the group stage is replaced by an incomplete round robin phase with 36 clubs that play against six or eight different opponents \citep{UEFA2023}. Incomplete round robin tournaments can also emerge from unevenly combining round robin tournaments between a subset of the opponents (e.g.\ teams in the same division) and bipartite round robin tournaments, where each team plays against each team from another group (e.g.\ teams in different conferences, or interleague play) as is common in North American sports.

%If not all rounds of a round robin tournament can be played -– for instance due to lack of time (especially if there are many players), because the tournament is aborted, or simply by design –- we call it an \emph{incomplete round robin tournament}. The consequence is that teams do not meet all opponents an equal number of times, which is why it is also called an \emph{unbalanced} design. This is common in North American sports that are organized in divisions and conferences (e.g.\ NBA), or the Australian Football League \cite{lenor2016}.

%The so-called \emph{incomplete round robin tournament} is also a non-eliminating tournament format where a predetermined number of rounds is played, which is fewer than $n-1$ \citep{Froncek2013}. It differs from the Swiss system tournament in that its pairings are determined before the beginning of the tournament, and, thus, do not depend on the outcomes of the matches.

\subsubsection{Knockout} \label{Sec212}

Another traditional competition format is the \emph{knockout tournament}. In contrast to round robin tournaments, here a player generally faces only a subset of the other players. In a single knockout (SKO) tournament, players are immediately eliminated if they lose.

In a double knockout (DKO) tournament, one loss is allowed in the tournament. This format is organized as two separate SKOs: the first SKO is called the main tournament, and the second SKO is called the consolation tournament. A player losing in the main tournament is moved to the consolation tournament. However, another loss in the consolation tournament eliminates the player. In essence, a player needs to lose twice in order to be knocked out. If the winner of the main round loses the final match against the winner of the consolation tournament, a rematch is played between them. The final phase of the Russian football cup is a double knockout tournament since the 2022/23 season.
A variant of this structure is used, for example, in judo, where the loser of the final in the main tournament fights against the winner of the consolation tournament for the second place. Thus, the winner will always have no losses, the silver medalist will have one loss, and the bronze medalist will have two losses.

In some knockout tournaments, the two players facing each other play more than one match. For instance, in the NBA playoffs, the opponents play as a ``best-of-7'': the team that first reaches four wins advances to the next round.
If multiple knockout tournaments are played by the same players, the SKO is said to be repeated \citep{david1959tournaments} or serial \citep{LambersPendavinghSpieksma2024}. The results of all subtournaments are aggregated to determine the final winner.

%\subsubsection{Seeding}

%\input{Figure1}

\begin{figure}[t]
\centering

\begin{subfigure}[b]{0.31\textwidth}
\centering
\caption{Balanced SKO with standard seeding}
\label{Fig1a}
\scalebox{0.5}{
    \begin{tikzpicture}[level/.style={sibling distance=(50mm/#1},rotate=180]
        \node[circle,draw] (root) {}
        child {
          node[circle,draw] {}
          child {
            node[circle,draw] {}
            child {node[circle, draw] {7}}
            child {node[circle,draw] {2}}
          }
          child {
            node[circle,draw] {}
            child {node[circle,draw] {6}}
            child {node[circle,draw] {3}}
          }
        }
        child {
          node[circle,draw] {}
          child {
            node[circle,draw] {}
            child {node[circle,draw] {5}}
            child {node[circle,draw] {4}}
          }
          child {
            node[circle,draw] {}
            child {node[circle,draw] {8}}
            child {node[circle,draw] {1}}
          }
        };
    \end{tikzpicture}
}
\end{subfigure}
\hspace{0.025\textwidth}
\begin{subfigure}[b]{0.31\textwidth}
\centering
\caption{Unbalanced SKO with an alternative seeding}
\label{Fig1b}
\scalebox{0.5}{
    \begin{tikzpicture}[level/.style={sibling distance=(50mm/#1},rotate=180]
  \node[circle,draw] (root) {}
        child {
          node[circle,draw] {}
          child {
            node[circle,draw] {}
            child {node[circle, draw] {5}}
            child {node[circle,draw] {3}}
          }
          child {
            node[circle,draw] {2}
          }
        }
        child {
          node[circle,draw] {}
          child {
            node[circle,draw] {1}
          }
          child {
            node[circle,draw] {}
            child {node[circle,draw] {6}}
            child {node[circle,draw] {4}}
          }
        };
    \end{tikzpicture}        
}
\end{subfigure}
\hspace{0.025\textwidth}
%\vspace{0.5cm}
\begin{subfigure}{0.31\textwidth}
\centering
\caption{Caterpillar}
\label{Fig1c}
\scalebox{0.5}{
    \begin{tikzpicture}[level/.style={sibling distance=(50mm/#1},rotate=180]
  \node[circle, draw](root){}
        child{
            node[circle, draw]{}
            child{
                node[circle,draw]{}
                child{
                   node[circle,draw]{}
                    child{ 
                        node[circle,draw]{}
                        child{
                            node[circle,draw]{6}
                        }
                        child{
                            node[circle, draw]{5}
                        }
                    }
                    child{
                        node[circle, draw]{4}
                    }
                }
                child{
                 node[circle, draw]{3}
                }
            }
            child{
                node[circle, draw]{2}
            }
        }
        child{
            node[circle, draw]{1}
        };
    \end{tikzpicture}        
}
\end{subfigure}
\caption{Different types of knockout tournaments}
\end{figure}

%In random knockout tournaments, the pairings in each round are done randomly. The study of random KO structures dates back to at least \citet{narayana1969contributions}. 

Knockout tournaments are characterized by their \emph{seeding} policy, which is typically used to avoid matches between the strongest contestants at the beginning of the competition. 
For instance, assume that eight participants are ranked from one to eight, with 1 being the strongest player, 2 being the second strongest player, etc. The \emph{standard} seeding for eight players is shown in Figure~\ref{Fig1a}. Notation-wise, we use brackets to represent the structure of the knockout tree, e.g.\ ((a,b),(c,d)) denotes that the winners of matches (a,b) and (c,d) face each other in the subsequent round. Thus, the standard seeding for eight players can be written as (((1,8),(4,5)),((2,7),(3,6))). The standard seeding ensures that in round $r=1,\dots,n$ of a tournament with $2^{n}$ players, the sum of ranks in every match of round $r$ is $2^{n-r+1}+1$ if always the favorite wins.
The tree diagram that represents the series of games to be played in a knockout tournament is called the \emph{bracket}. 
If the seeding is done randomly, the knockout is said to be \emph{unseeded} or random; the study of random KO structures dates back to at least \citet{narayana1969contributions}.

%Since the seeding determines the order of the matches a specific team has to play, we discuss this in a seperate section (\ref{sec:seeding}).

%\subsubsection{Balanced and unbalanced knockout tournaments}

In a \emph{balanced} SKO, each player should win the same number of matches in order to win the tournament. This is not the case for an \emph{unbalanced} SKO: in Figure~\ref{Fig1b}, players 1 and 2 enter the tournament later than the other players and can claim the tournament victory by winning one game less. The most unbalanced case is shown in Figure~\ref{Fig1c}, which is known as a ``caterpillar''.
Unbalanced knockout tournaments are organized in a variety of sports, for instance, in baseball, boxing, bowling, and squash. A prominent example is the qualification for UEFA club competitions \citep{Csato2022b}.

\subsubsection{Swiss system} \label{Sec213}

The \emph{Swiss system} is a non-eliminating tournament format where a predetermined number of rounds is played. Contrary to the round robin format, it contains fewer rounds than the $n-1$ required to meet all players in a tournament with $n$ players. Its key characteristic is that opponents are determined dynamically. In the first round, the matches depend on an exogenous ranking of the players. In all subsequent rounds, opponents are matched based on several criteria, but the primary goal is typically to play against an opponent having approximately the same number of points \citep{SauerCsehLenzner2024}. The Swiss system was first used at a chess tournament in Z\"urich in 1895, but its variants have recently been applied to schedule other sports competitions, including e-sports tournaments \citep{dong2023dynamic}.

%The so-called \emph{incomplete round robin tournament} is also a non-eliminating tournament format where a predetermined number of rounds is played, which is fewer than $n-1$ \citep{Froncek2013}. It differs from the Swiss system tournament in that its pairings are determined before the beginning of the tournament, and, thus, do not depend on the outcomes of the matches.

%The Swiss system raises particularly interesting research problems regarding the matching algorithm \citep{Olafsson1990, kujansuu1999stable, BiroFleinerPalincza2017, SauerCsehLenzner2024}, since this is---in contrast to the round robin and knockout systems---far from obvious.

% The Swiss system fixes the number of rounds in advance. Characteristically for the Swiss system is that often, there are more rounds than players. As a result, a team faces only a subset of its possible opponents. Similarly, a k-incomplete round robin (k-iRR) is scheduled at the beginning of the season but with the restriction that each team can face each other team at most (instead of exactly) k-times. The analysis of incomplete round robin tournaments dates back to at least \citet{cochran1971designs}. They have received serious attention after the COVID-19 crisis, especially with regards to constructing a fair ranking \citep{lambers2020true, csato2021coronavirus, gorgi2023estimation, van2023probabilistic}. During this crisis, almost all sports competitions were suddenly shut down, accidentally turning most round robin tournaments into incomplete tournaments.

\subsubsection{Multi-stage and hybrid tournaments} \label{Sec214}

%A hybrid tournament consisting of a SRR group stage followed by a SKO stage is an example of a 2-stage tournament. 
Tournaments can be played in multiple stages. Typically, only a subset of the players is allowed to proceed to the next stage, based on their performance in the previous stage. If the tournament formats differ in at least two stages, the tournament is commonly called \emph{hybrid}. The FIFA World Cup, with a SRR group stage followed by a knockout stage for the best teams, is a well-known example. 

\subsection{Ranking} \label{Sec22}

The ultimate aim of organizing a tournament is to rank the contestants. Section~\ref{Sec221} introduces ranking in an individual tournament, Section~\ref{Sec222} deals with the aggregation of rankings, and Section~\ref{Sec223} gives a concise overview of global rankings, which rank the players on the basis of several tournaments.

\subsubsection{Tournament ranking and tie-breaking} \label{Sec221}

A tournament ranking gives an order of the teams based on their achievements in the tournament.
%In a multi-stage tournament, the seeding of the teams in one stage is usually determined by the ranking in the previous stage.
In some sports, performance can be directly measured by distances or finishing times. In the case of paired comparisons, the score for each possible match outcome (win or loss, in some cases tie) needs to be decided. In football, for instance, a win is typically worth 3 points, a draw 1 point, and a loss 0 points, which is the (3,1,0) scoring/ranking system.

While various sophisticated ranking methods exist, the most widely used approach is to sum these scores. Naturally, this method does not take the quality of a win into account, i.e.\ matches played against a strong and a weak opponent yield the same number of points. This can be a problem especially in incomplete round robin and Swiss system  tournaments, as will be discussed in Sections~\ref{Sec721} and \ref{Sec723}, respectively.

Summing the scores resembles counting the number of opponents a player has beaten, which is the only ranking rule satisfying three appealing axioms \citep{rubinstein1980ranking}: anonymity (the ranking method does not depend on the identity of the players), positive responsiveness (a loss does not result in a better rank than a win), and autonomous relative ranking (the relative ranking of two players is not influenced by a match where neither of them is involved in). \citet{henriet1985copeland} extends this characterization by allowing for ties, which are worth half of a win. While this method is used in some team sports where ties are allowed such as handball, the (3,1,0) scoring system has no theoretical background according to our knowledge.

Another well-known property from social choice theory is \emph{inversion} \citep{ChebotarevShamis1998a}: the ranking is reversed if all results are reversed. Consider a SRR with four players $i$, $j$, $k$, $\ell$ where $i$ won against $j$ and $\ell$, $j$ won against $\ell$, while the three matches of $k$ resulted in a draw. This example shows that  the (3,1,0) scoring system does not satisfy inversion. Indeed, using the notation ($=$, $>$) for the pairwise majority relation based on the aggregated score, the ranking according to the (3,1,0) method is $i > j > k > \ell$, and the ranking after all results are reversed is $\ell > j > k > i$. Naturally, the (2,1,0) scoring system satisfies inversion; in our example, the original ranking is $i > j = k > \ell$, and the ranking after all results are reversed is $\ell > k = j > i$.

Often a complete ranking of the players is desired. Therefore, ranking methods contain various tie-breaking rules to rank players with the same score. In round robin tournaments, the most popular criteria are the number of matches won, head-to-head results, and superior goal difference \citep[Chapter~1.3]{Csato2021a}. The effects of choosing a specific hierarchy of tie-breaking rules are often non-negligible; for example, they influence the probability of \emph{heteronomous} relative ranking, i.e.\ the relative ranking of two teams depends on the outcome of a match in which neither team was involved \citep{berker2014tie}.
%\citet{pakaslahti2019use} discusses the negative consequences of including head-to-head results and argues that in round robin tournaments, goal difference should be preferred over head-to-head records when breaking ties.

Knockout tournaments also require tie-breaking rules in some sports such as football and handball. Tied matches are usually followed by extra time, and the final tie-breaking is usually a penalty shootout since the 1970s \citep{Palacios-Huerta2014}. UEFA introduced the away goals rule in the 1965/66 European Cup Winners’ Cup for two-legged clashes. According to this principle, if both teams score the same number of goals in the two games, the team that scores more goals away will advance.

%Swiss system tournaments use several tie-breaking rules (see \url{https://en.wikipedia.org/wiki/Tie-breaking_in_Swiss-system_tournaments} for a concise overview). 
The majority of tie-breaking rules used in Swiss-system tournaments are based on the idea that players playing against stronger opponents should be preferred. One of the most extensively used methods is \emph{Buchholz}, which is essentially the arithmetic mean of the scores obtained by the opponents. \citet{Freixas2022} suggests using a weighted average of the opponents’ scores where the weights are assigned according to the binomial distribution.

\subsubsection{Aggregation of rankings} \label{Sec222}

%\begin{table}[t]
%    \centering
%    \caption{The Formula One points scoring system since 2019}
%    \label{Table1}
%\begin{threeparttable}
%    \begin{tabularx}{0.8\textwidth}{l CCCCC CCCCC c} \toprule
%       Position & 1 & 2 & 3 & 4 & 5 & 6 & 7 & 8 & 9 & 10 & Bonus \\  \midrule
%       Points & 25 & 18 & 15 & 12 & 10 & 8 & 6 & 4 & 2 & 1 & 1 \\ \bottomrule
%    \end{tabularx}
%    \begin{tablenotes} \footnotesize
%        \item
%        Bonus points are awarded for the fastest lap if the driver finishes in the top 10 positions.
%    \end{tablenotes}    
%\end{threeparttable}
%\end{table}

Some tournaments consist of individual contests, which have their own rankings. Examples include cycling (e.g.\ Tour de France), motorsport racing (e.g.\ Formula One), or winter sports (e.g.\ Biathlon World Cup). Here, a score is typically associated with each position in each contest, and they are added to obtain the final tournament ranking. %As an example, Table~\ref{Table1} shows points scoring system of Formula One used since 2019.
A similar aggregation of rankings can be favourable even if the official rule is based on the ``objective'' finishing times: \citet{Ausloos2024} argues for ranking teams in cycling races by the cumulative sums of the places of their riders instead of their finishing times.
However, the aggregation can lead to different paradoxes; \citet{Kaiser2019} presents some of them from Formula One. Since rank aggregation is an extensively studied problem in social choice theory, we do not go into details; a recent summary is provided by \citet{KondratevIanovskiNesterov2023}.

\subsubsection{Global rankings} \label{Sec223}

Ranking can also refer to the global ranking of players in a particular discipline. In that case, the ranking is often based on a rating, which quantifies their strength relative to each other. Since players in this case typically did not face equally strong opponents, the rating should account for the strength of their opponents.
The widely used rating method developed by Arpad Elo, a Hungarian-born physics professor, does exactly that \citep{elo1978rating, Aldous2017}. After each match, the rating of a player is updated by the formula
\begin{equation}
    r_{new} = r_{old} + K(S- \mu),
\end{equation}
where $K$ is a sports-specific parameter reflecting the maximum possible adjustment per game, $S$ is the outcome of the match (typically 1 for a win and 0 for a loss), and $\mu$ represents the expected outcome. The latter is determined by the ratings of players $r_i$ and $r_j$, as well as the scaling parameter $s$ according to the following formula:
\begin{equation}
    \mu = \frac{1}{1+10^{(r_{j}-r_{i})/s}}.
\end{equation}
Naturally, the Elo method can be parameterized in different ways depending on the value of $K$, the definition of match outcome $S$, or the adjustment of $r_i$ due to the potential home advantage. \citet{GomesdePinhoZancoSzczecinskiKuhnSeara2024} present a comprehensive stochastic analysis of the Elo algorithm in round robin tournaments, and investigate the effect of parameters $K$ and $\mu$ on its performance to obtain design guidelines.

FIFA, for instance, after having received serious criticism in the academic literature \citep{lasek2016improve, cea2020analytics, Csato2021a, Kaminski2022}, adopted the Elo method for their world ranking \citep{FIFA2018c}.
The popularity of Elo ratings is shown by the fact that the FIFA world ranking has an alternative, the World Football Elo Ratings (\url{https://www.eloratings.net/}), that takes home advantage and goal difference into account as suggested by \citet{SzczecinskiRoatis2022}. Analogously, the performance of football clubs is measured by at least two independent websites, \url{http://clubelo.com/} and \url{http://elofootball.com/}.
Nonetheless, some global rankings do not take the strength of opponents into account. For example, UEFA club coefficients are based on awarding points for wins and reaching certain stages in previous UEFA club competitions \citep{DagaevRudyak2019}.

%Nevertheless, the construction of the global ranking can have a major impact on individual tournaments. In tennis, only the 104 highest-ranked players qualify automatically for the prestigious Grand Slam tournaments. Seeding is also determined by a global ranking if no previous stage has been played in a tournament.
%For instance, in 1992, FIFA started to publish a rating of its member nations so that the relative strengths of football nations could be compared. 

There are many sports-specific ranking rules and rating systems.
\citet{stefani2011methodology} lists the official rating systems of 159 sports federations, while \citet{LangvilleMeyer2012} provide an overview of the mathematics of various ranking methods. An overview of ranking systems in football can be found in \citet{vanEetveldeLey2019}.

\subsection{Design criteria} \label{Sec23}

Sections~\ref{Sec231}--\ref{Sec233} shortly discuss the three main criteria that have been identified to evaluate a tournament design: efficacy, fairness, and attractiveness.

\subsubsection{Efficacy} \label{Sec231}

The ability of a tournament to rank contestants according to their true strength is called \emph{efficacy} \citep{lasek2018efficacy, Sziklai2022}. Efficacy with respect to only the best player is also known as \emph{effectivity} \citep{glenn1960comparison} or \emph{predictive power} \citep{ryvkin2008predictive, vu2011fair}.

By far the most popular way of evaluating these properties is to use simulation: all matches of the tournament are simulated and the observed ranking is compared with the assumed true ranking of players.By aggregating the results over many simulation runs (i.e.\ Monte Carlo simulation), meaningful insights can be derived about how efficacious the tournament is.
Efficacy is mostly studied from the perspective of the tournament format and the seeding rule.

\subsubsection{Fairness} \label{Sec232}

In contrast to efficacy, there is no uniform definition of fairness as there exists no universally agreed system of ethics \citep{Baumer2024, Palacios-Huerta2023}. Still, we have identified two main principles of fairness. The first one is based on the idea that, \emph{a priori}, a player should not be disadvantaged compared to any weaker player. This means that, for every player, the probability of winning the tournament should be higher than that of any weaker player. This concept is also commonly called \emph{order preservation} \citep{Schwenk2000, Groh2008, vu2011fair, prince2013designing, karpov2016new, placek2023impossibility}.
%In order to satisfy this principle, some authors advocate that strong players should be favored over weak players. In a knockout tournament, this can be accomplished by matching stronger teams with weaker teams in the first round. However, this means that the strongest players have the easiest routes to the final. In some knockout tournaments, some teams enter the competition later, meaning they have to win fewer matches in order to become the winner \citep{stanton2013structure}.

The second approach of fairness takes on a completely different perspective. It is based on the principle of treating all players equally and tries to eliminate any favoritism towards particular players. For instance, if players are allocated to groups based on a draw, the procedure is said to be fair if all allocations are equally likely. Analogously, fairness in scheduling is mostly related to this principle by balancing some objectives (the number of rest days, the distances to be traveled, the number of consecutive home or away games, etc.) over the teams. Equal treatment of all players is sometimes weakened to equal treatment of equally strong players \citep{Guyon2018a, Arlegi2020, Csato2022d}. 

Note that the first fairness principle is closely related to efficacy. In fact, the main reason why an organizer would want to adhere to this principle is to guarantee efficacy. Hence, to reduce the overlap between the discussion of our design criteria, fairness will mean the principle of equal treatment in the following. Therefore, papers studying fairness from the perspective of order preservation will be presented in the sections dealing with efficacy.  

\subsubsection{Attractiveness} \label{Sec233}

The attractiveness of a match can be defined in multiple ways: by its quality, competitive intensity, and importance.
The quality of a match played by players $i$ and $j$ with their strengths $r_i$ and $r_j$, respectively, can be quantified as $r_i+r_j$, while its competitive intensity can be defined as $- \lvert r_i-r_j \rvert$ \citep{Csato2020b, DagaevRudyak2019, Dagaev2018}. Both measures seem to be linked with higher attendance rates \citep{forrest2002outcome, garcia2002determinants} and TV audiences \citep{schreyer2018game}.

On the other hand, the importance of a match is usually defined with respect to a certain prize. \citet{jennett1984attendances} suggested a simple measure to estimate the uncertainty of winning the championship \emph{ex post}. Let $k$ be the total number of points that turned out to be necessary to win the championship. The importance of a match is the inverse of the minimum number of matches required to achieve $k$ points. This metric was later used by \citet{borland1992attendance} and \citet{dobson1992demand} to investigate attendance rates.
A decade later, \citet{schilling1994importance} introduced probably the importance measure that is probably most well-known. The Schilling importance $_{i}S(X)_{t,t+k}$ of match $t+k$ at time period $t$ for team $i$ is defined as:
\begin{equation}
    _{i}S(X)_{t,t+k} = Pr(_{i}X|_{i}W_{t+k}, H_t) - Pr(_{i}X|_{i}L_{t+k}, H_t),
\end{equation}
where $Pr(_{i}X|_{i}W_{t+k}(L_{t+k}), H_t)$ is the probability that team $i$ wins prize $X$ (e.g.\ avoiding relegation) if $i$ wins (loses) the match $t+k$, given the history $H_{t}$ of all matches played before $t$. \citet{Lahvicka2015a} extends Schilling's measure to account for ties. 

Many other ways have been suggested to quantify match importance. \citet{GoossensBelienSpieksma2012} take into account the number of teams that are still competing for a given prize. An integer program is used to determine the highest and lowest rank any team can still obtain at the end of the season when a certain match is played. Similar formulations are proposed by \citet{ribeiro2005application, raack2014standings, GotzesHoppmann2022}. \citet{Geenens2014} measures the decisiveness of a match based on the expected entropy (i.e.\ uncertainty). \citet{corona2017importance} use an entropy-based measure to identify the most decisive matches in the 2014 FIFA World Cup and the 2015 Copa Am\'erica. \citet{GollerHeininger2024} extend the entropy measure by considering the difference between the reward probability distributions derived from the possible outcomes of a single event. \citet{CsatoMolontayPinter2024} introduce a classification scheme to determine weakly (where the final rank of one team is known, regardless of the match outcome) and strongly (where the match outcome does not influence the ranking of both teams) stakeless matches. 

There is a wide consensus in the literature that still being in contention for a prize positively affects attendance \citep{GarciaRodriguez2009, pawlowski2015competition, losak2024does}. In particular, the tightness of the contention seems to control the size of this effect \citep{krautmann2011playoff}.
Furthermore, \citet{BuraimoForrest2022} show that match importance is the key driver for the size of TV audiences in the case of English football games instead of the much more researched outcome uncertainty (see the so-called outcome uncertainty hypothesis \citep{rottenberg1956baseball}, and \citet{SchreyerAnsari2022} and \citet{vanderBurg2023} for recent overviews).

%Whether match-level uncertainty of outcome indeed leads to higher stadium attendance or TV ratings is a much researched topic in economics (see e.g.\ \cite{pawlowski2013} and \cite{ vanderBurg2023} for an overview on soccer). 

While importance is usually measured at the level of individual games, suspense refers to uncertainty in the outcome of the tournament. Although suspense can be related to any outcome, such as relegation or qualification for international competitions, it is typically interpreted within the context of the tournament winner. Suspense is an important measure of attractiveness as it is positively related to both stadium attendance \citep{pawlowski2015competition} and TV audience \citep{bizzozero2016importance, buraimo2020unscripted, BuraimoForrest2022}.

\section{Tournament format} \label{Sec3}

The comparison of tournament formats has received much attention in operational research since determining the format is probably the first task of the contest organizer. Even if the number of games that can be played is strictly limited, usually several options remain available. For example, \citet{Guyon2020a} uncovers a serious drawback of the original format of the 2026 FIFA World Cup and lists 7 alternative options. \citet{Renno-Costa2023} presents another one based on double elimination, while \citet{GuajardoKrumer2024} develop three further formats.
Choosing the best from such a set of feasible designs clearly requires their evaluation with respect to a number of criteria.

\subsection{Efficacy} \label{Sec31} 

At its inception, the tournament design literature was mainly concerned with the probability of the best player winning under a given tournament format, also called effectivity. Pioneering work was done by \citet{david1959tournaments}, who used combinatorial techniques to compute exact winning probabilities for a SKO and SRR. To that end, a series of pairwise win probabilities needs to be assumed; $\pi_{ij}$ is the probability that $i$ beats $j$, which are collected in matrix $\Pi$. \citet{david1959tournaments} assumes a cumulative normal distribution $\Phi(\cdot)$ for the pairwise win probabilities, i.e.\ $\pi_{ij} = \Phi\{\frac{1}{\sigma}(r_{i} - r_{j})\}$, with $\sigma$ being the variance term and $r_i$ being the ``true'' rating of player $i$. \citet{glenn1960comparison} and \citet{searls1963probability} apply a similar approach to evaluate tournament structures, although they assume an arbitrary matrix $\Pi$.
%\citet{searls1963probability} considers a multistage tournament organized as a SKO in both stages with eight players. If the winner of the first stage loses in the second stage, a final match between the winners of the two stages determines the ultimate winner. %This design will be denoted by SKO+SKO$^{\text{playoff}}$. 
All three papers compute the probabilities that other players finish in a certain position as well. 
Although limited to tournaments with 8 players, the authors find SKO to be the least effective format. Double and serial knockout formats are more effective, in particular, if matches are played as a best of three.
%\cite{david1959tournaments} investigates some properties of KO and RR competitions and develops expressions to compute the probability with which the best player wins the tournament, when certain assumptions about the strength of players can be made. \cite{glenn1960comparison} also develops expressions for various 4-player tournament designs such as a double elimination tournament and a single KO where each match-up between two players is designed as a best of three contest. In order to compute exact probabilities, a series of pairwise win probabilities is assumed ($\pi_{ij}$ being the probability that $i$ beats $j$). The standard single KO was found to be the least effective, while the best of three single KO was found to be the most effective. The SRR was found to be more effective than the standard single KO, with the double, triple and double elimination KO all being even slightly better. In line with this study, \cite{searls1963probability} defines similar expressions for tournaments with eight players. The double elimination tournament was extended to comprise best of three series. This design was found to be the most effective among the standard single, best of three KO and the standard double elimination tournament. Moreover, \cite{david1959tournaments}, \cite{glenn1960comparison} and \cite{searls1963probability} all developed expressions to evaluate the probabilities that other players are ranked in a certain position as well. 
For random knockout tournaments, \citet{KnuthLossers1987} compute the chance of player $i$ winning the tournament, assuming that it wins against player $i+1$ with probability $p$ and always defeats player $j$ if $j > i+1$. 
\citet{adler2017random} derive an upper and lower bound on the probability for of winning the tournament for all players, if player $i$ wins against player $j$ with probability $v_i / \left( v_i + v_j \right)$, where the values $v_1$, \dots , $v_n$ are exogenously given.

As the exact computation of these expressions quickly becomes intractable, the standard approach in the literature is to perform Monte Carlo simulation. 
Using this approach, several papers conclude that round robin tournaments are typically more efficacious than knockout tournaments \citep{Appleton1995, mcgarry1997efficacy, ryvkin2008predictive, Scarf2009, ryvkin2010selection, scarf2011numerical, Csato2021b}. Furthermore, the efficacy of a SKO increases by extending it towards a DKO \citep{mcgarry1997efficacy, stanton2013structure}.
Efficacy can usually be improved by increasing the number of matches \citep{lasek2018efficacy} but the relationship is not necessarily monotonic for complex hybrid tournaments \citep{Csato2021b}. According to earlier studies \citep{Appleton1995, dinh2020simulating}, the Swiss system is rather weak in identifying the true ranking of players. However, \citet{Sziklai2022} find that the Swiss system outperforms other formats with the same number of games. To the best of our knowledge, the efficacy of incomplete round robin tournaments has not been examined yet. 

\citet{Lahvicka2015b} studies the effect of the additional playoff stage on the efficacy of a DRR in the Czech ice hockey league. The average probability that the strongest team becomes a champion is found to decrease from 48\% to 39\%. The reduction is even higher if the strongest team is more dominant, and is robust with respect to relaxing the assumptions of constant team strengths and the lack of strategic behavior. This is an important message to administrators: uncertainty can be increased by organizing an extra knockout phase after a round robin tournament.

%Furthermore, \citet{Sziklai2022} conclude that seeding only has a marginal effect in practice because it is a) difficult to estimate the true ranking of participants before the beginning of the competition; and b) the actual performance can substantially differ from past performance, making the seeding unreliable. The effect of the actual seeding method will be discussed in the next section.

%\input{Figure2}

The balancedness of a knockout tournament also affects its efficacy. Balanced knockout tournaments with eight or more players cannot satisfy simultaneously that a) stronger players are favored over weaker players; and b) equally strong players are treated equally \citep{vu2011fair, Arlegi2020}. In particular, pairwise probability matrices can be found to show the incompatibility of these two axioms, see \citet{kulhanek2020surprises} and \citet{arlegi2022can}.
Stronger players can be favored more explicitly by an unbalanced knockout tree. In the most extreme case of a caterpillar SKO (Figure~\ref{Fig1c}), a stronger player can always be favored over a weaker player \citep{vu2011fair}. \citet{Arlegi2020} generalize this finding: any unbalanced tree is verified to be efficacious if it satisfies the following two properties:
a) it does not contain any \emph{antler} (a special substructure, see \citet{Arlegi2020}); and
b) the seeding rule places weaker players to higher levels in the knockout tree than stronger players.

\begin{table}[t!]
    \centering
    \caption{Studies on the efficacy of tournament formats}
    \label{Table1}

\begin{subtable}{\textwidth}
    \centering
    \caption{Papers focusing on efficacy}
    \label{Table1a}
\rowcolors{1}{}{gray!20} 
    \begin{tabularx}{\textwidth}{LL cccc} \toprule
        Study & Format & $\pi_{ij}$ & Best & Worst & $n$ \\ \bottomrule
        \citet{mcgarry1997efficacy} & \Centerstack[l]{SKO, SKO$_\text{S}$, SKO$_\text{A}$, \\ DKO, DKO$_\text{S}$, SRR, \\ SRR+SKO, SRR+SRR} & $\Pi^{a}$ & \Centerstack{DKO$_\text{S}$, \\ SRR, \\ SKO$_\text{S}$} & SKO$_\text{a}$ & 8 \\
        \citet{Scarf2009} & \Centerstack[l]{DRR, DKO, DKO$_\text{S}$, \\ SKO$_\text{S}$, SRR+DKO, \\ SRR+DKO$_{\text{S}}$, DRR+DKO, \\ DRR+DKO$_{\text{S}}$, DRR+SKO$_{\text{S}}$} & Poisson$(\cdot)$ & DRR & SKO & 32 \\
        \citet{scarf2011numerical} & 23 formats & Poisson$(\cdot)$ & SRR & SKO  & 32\\
        \citet{stanton2013structure} & SKO, DKO & CRM, $\frac{1}{2} + \frac{j-i}{2(N-1)}$ & DKO & SKO & 8--$10^{24}$ \\
        \citet{lasek2018efficacy} & \Centerstack[l]{SRR, multistage RR} & Poisson$(\cdot)$ & 3RR + (SRR/SRR) & SRR & 12, 16, 18 \\
        \citet{dinh2020simulating} & \Centerstack[l]{SKO, DKO, SKO$_{\text{S}}$, \\ SRR, Swiss, Reaper T} & NS & SRR & SKO & 8 \\
        \citet{Csato2020b} & \Centerstack[l]{Three hybrid formats: \\ D(8+6), D(4$ \times 7$), \\ D(4$ \times 6$)$^1$} & $\frac{(57-r_{i})^{k}}{(57-r_{i})^{k} + (57-r_{j})^{k}}$ & D(8+6) & D($4 \times 6$) & 24, 28 \\
        \citet{Csato2021b} & \Centerstack[l]{Four hybrid formats: \\ SRR, SKO, G64, \\ G66, G64$^2$} & $\frac{1}{1 + \left[ \left( r_i + 24 \right) / \left( r_j + 24 \right) \right]}$ & G66 & G64 & 24 \\
        \citet{Sziklai2022} & \Centerstack[l]{SRR, DRR, SKO, \\ SKO-3, DKO, \\ SRR+SKO, Swiss} & $\frac{1}{1+10^{(r_j-r_i)/400}}$ & Swiss & SKO & 32  \\ \toprule
    \end{tabularx}
\end{subtable}

\vspace{0.5cm}
\begin{subtable}{\textwidth}
    \centering
    \caption{Papers focusing on effectivity}
    \label{Table1b}
\begin{adjustbox}{width=\textwidth}
\begin{threeparttable}
\rowcolors{1}{}{gray!20} 
    \begin{tabularx}{\textwidth}{lLC ccc} \toprule
        Study & Format & $\pi_{ij}$ & Best & Worst & $n$ \\ \bottomrule
        \citet{david1959tournaments} & \Centerstack[l]{SRR, SKO, \\ serial SKO} & $\Phi\{ \left( r_{i} - r_{j} \right) / \sigma \}$ & serial SKO & NS & 3-8 \\
        \citet{glenn1960comparison} & \Centerstack[l]{SRR, SKO, SKO-3, \\ serial SKO, DKO} & $\Pi^{a}$ & SKO-3 & SKO & 4 \\
        \citet{searls1963probability} & \Centerstack[l]{SKO, SKO-3, \\ SKO+SKO, \\ SKO-3+SKO-3} & $\Pi^{a}$ & SKO-3+SKO-3 & SKO & 8 \\
        \citet{Appleton1995} & \Centerstack[l]{SKO, SKO$_\text{S}$, DKO, DKO$_\text{S}$, \\ SKO-3, SRR, DRR, \\ SRR-SKO, Swiss} & $\Phi\{k(r_{i} - r_{j})\}$ & DKO$_S$, DRR & SKO & 8, 16 \\
        \citet{ryvkin2008predictive} & SKO, SRR & pdf$^3$ & SRR & SKO & NS\\
        \citet{ryvkin2010selection} & SKO, SRR & pdf$^3$ & SRR & SKO & NS\\ \bottomrule
    \end{tabularx}
    \begin{tablenotes} \footnotesize       
        \item NS: not specified
        \item $k$RR+$x$KO stands for a hybrid format where $k$RR is followed by $x$KO.
        \item The subscript S as in SKO$_{S(A)}$ indicates that the SKO is seeded according to the standard (an alternative) seeding.
        \item $x$KO-$y$: best of $y$ knockout format
        \item[1] D(8+6): 2 groups of 8, 2 groups of 6; D($g$, $n$): $g$ groups of $n$ teams, all followed by a knockout stage.
        \item[2] G$xy$: stage 1 with groups of $x$ teams and stage 2 with groups of $y$ teams, followed by a knockout stage.
        \item[3] Due to the notational complexity of probability density functions (pdf), we refer to the paper itself for the formulas for $\pi_{ij}$.
    \end{tablenotes}
\end{threeparttable}
\end{adjustbox}
\end{subtable}

\end{table}

A structured overview of the papers that either compute or simulate winning probabilities is presented in Tables~\ref{Table1}. The formats considered are given in the second column. The third column shows the pairwise win probabilities $\pi_{ij}$ that were used to compare the formats. Using Poisson distribution or arbitrary win probabilities turns out to be the most popular approaches. Next, the best and worst formats of all formats studied are reported. Finally, the last column indicates the number of players in the tournaments compared.
%As can be seen from table \ref{Table3}, round robin tournaments are usually found more efficacious than knockout tournaments. In general, if each pair of contestant plays more games, the tournament becomes more efficacious. However, this obviously contains a trade-off, reflected by the number of rounds and matches required. Since a player is immediately eliminated after one loss in a SKO, upsets have a powerful effect on the final ranking compared to a SRR. Unsurprisingly, most of the literature argues that the efficacy of SKOs can be increased by extending it to a format where the players are not eliminated after one loss, such as in a double elimination or best-of-three knockout. Analogously, the efficacy of round robin tournaments will also improve by extending the SRR to a DRR or even a 3RR. 

As can be seen from Table~\ref{Table1}, almost all studies use different statistical models, each with its limitations. This raises questions about whether the results can be attributed to the tournament format or to the model assumptions. However, as noted by \citet{Appleton1995}, these simulation studies primarily aim to compare different tournament formats, and not to correctly estimate the chance of a player winning any particular tournament. Still, it would be interesting to see how the efficacy of tournament formats is influenced by the assumptions that are made about the strengths of the teams. 

Finally, the effect of the number of participants ($n$) on efficacy is still a blind spot. \citet{MullerHoodSokol2023} use simulations to analyze how expanding the NCAA college football championship from four to eight or 12 teams affects the probability that the championship tournament includes and is won by the true best team.
%They found that the effect of this expansion is almost negligible. However, with 12 teams the championship tournament is 2-12\% more likely to include the true best team, but it would be 5-6\% less likely to win the championship.
Further research in this direction would be particularly welcome since there exist many examples of well-known tournaments that have seen a change in the number of participants: the FIFA World Cup is organized with 32 teams in 2022 but 48 in 2026, the UEFA European Championship was contested by 16 teams in 2012 but by 24 in 2016, and the World Men's Handball Championship expanded from 24 to 32 teams between 2017 and 2019.

\subsection{Fairness} \label{Sec32}

The principle of equal treatment is discussed separately for the main tournament formats.

\subsubsection{Round robin tournaments} \label{Sec321}

In a round robin tournament, no player is a priori advantaged over any other player since each player faces each other player the same number of times. In the case of a $k$RR with odd $k$ and matches played at the venue of the home team, a possible element of unfairness can be that different teams enjoy their home advantage against different teams. This happened for instance in the UEFA Cup between 2004 and 2009, when the group stage had groups of 5 playing SRR. Additionally, if the number of teams is even, half of the teams have one more home game than the remaining teams. This is an issue, for example, in the Faroe Islands and Hungarian domestic football league, which are 3RR tournaments. 

In an incomplete round robin tournament, teams do not face each opponent the same number of times, which conflicts with the principle of equal treatment. One solution can be to determine each team’s opponents such that the total strength of the opponents is the same over all teams. \citet{froncek2006} and \citet{Froncek2007} give conditions for the existence of this so-called \emph{fair incomplete tournament}. On the other hand, \citet{Froncek2013} designs an incomplete round robin tournament where all teams have \emph{a priori} roughly the same chance of winning since the highest-ranked teams face the toughest opponents, and weaker teams play against weaker opponents; he calls this a handicap incomplete tournament. 

Sporadically, unforeseen circumstances (e.g.\ COVID-19) force a round robin tournament to be incomplete. \citet{HassanzadehHosseiniTurner2024} focus on the problem of finishing a suspended league where there are not enough rounds left to complete all matches. They propose a data-driven model that exploits predictive and prescriptive analytics to select a subset of originally scheduled games, in a way that the resulting ranking is similar to the ranking that would have resulted had the season been completed fully.

\subsubsection{Knockout tournaments} \label{Sec322}

Fairness is an important criterion in serial knockout tournaments such as the format that has recently been adopted by the Professional Darts Corporation. According to \citet{LambersPendavinghSpieksma2024}, each pair of players should potentially meet equally often in each round. By making a connection to Galois fields, they construct a stable tournament with $n-1$ subtournaments for any number of players $n$ that is a power of two.

Contrary to a serial SKO, the four prestigious Grand Slam tennis tournaments are SKOs played independently from each other. However, since the seeding only aims to avoid matches between top players in earlier rounds, a weak player sometimes has to face strong players over-abundantly.
%The goal of seeding is typically to avoid matches between top players in earlier rounds, rather than ensuring equal treatment of players.
Therefore, \citet{DellaCroceDragottoScatamacchia2022} develop an integer quadratic program together with a heuristic to maximize the diversification of pairings in the first round. Furthermore, repeated pairings over the tournaments are limited as much as possible. 

\subsubsection{Swiss system tournaments} \label{Sec323}

In these tournaments, the pairing for a round depends on the results of the previous rounds (see Section~\ref{Sec213}). Therefore, the tournament format can be described only together with the pairing rules that determine the set of matches to be played in the next round.
The main goals are as follows:
(1) players with (approximately) equal scores should play against each other;
(2) each player alternately has the advantage (e.g.\  play white in chess or play at home in other sports) to the extent possible; and
(3) no player is allowed to face the same opponent twice.

\citet{Olafsson1990} reveals how the pairing can be implemented by solving a weighted matching problem, which converts the pairing rules into penalty points.
Similarly, \citet{BiroFleinerPalincza2017} show that the more complex matching procedures as described by the World Chess Federation (FIDE) can be calculated in polynomial time with maximum weight matching.
\citet{kujansuu1999stable} suggest using the well-known stable roommates algorithm to determine the pairs in a Swiss system tournament such that each player has a preference list on the other players, where the preferences are constructed to reflect the criteria above.
\citet{SauerCsehLenzner2024} present an innovative pairing rule by computing maximum weight matchings in a carefully designed graph. On the basis of extensive simulations, the proposed procedure yields fairer pairings, and produces a final ranking that reflects the players’ true strengths better than the FIDE pairing system.

%To summarise, fairness in round robin tournaments has mostly been studied from the perspective of scheduling as will be presented in Section~\ref{Sec92}. Similarly, fairness in knockout tournaments has mostly been studied from the perspective of efficacy as we have seen in Section~\ref{Sec51}. Hence, fairness in tournament formats is not discussed further here.

\subsection{Attractiveness} \label{Sec33}

In some tournament formats, unimportant matches are difficult or even impossible to avoid. \citet{faella2021irrelevant} define a match as irrelevant if its outcome cannot change the ranking, and prove that $k$RR tournaments with more than five players, a fixed schedule and ranking by the win-loss method may contain irrelevant matches. %However, dynamic schedules for an arbitrary number of players can be devised to eliminate irrelevant matches for at least one of the players involved in each match.
\citet{scarf2008importance} calculate the Schilling importance and conclude that 52\% of the matches in the second half of the 2004/05 English Premier League season did not have any effect on which team would win the league. Hence, although RRs are typically quite efficacious, they are also characterized by a high proportion of unimportant matches. According to \citet{Scarf2009}, on average 18\% of all matches are unimportant in a DRR.
%Contrarily, in a SKO every match can be deemed important, since a loss directly eliminates the player.  

\citet{GoossensBelienSpieksma2012} simulate match results based on historical data to compare four different competition formats considered for the Belgian football league. The 2RR format used at that time scored rather poorly; indeed, the tournament format of the league was reformed not much later. \citet{Csato2020b} compares the expected quality and competitive balance of alternative tournament formats for the EHF Men's Handball Champions League. Similarly to \citet{GoossensBelienSpieksma2012}, formats other than the traditional one are revealed to increase the attractiveness of the tournament.

Stakeless games (see Section~\ref{Sec233}) are detrimental not only due to problems with incentives but they are also probably much less interesting for the audience. The FIFA World Cup format has been extensively studied from this perspective \citep{ChaterArrondelGayantLaslier2021, Guyon2020a, Stronka2024}.
The novel incomplete round robin format of the UEFA Champions League (Section~\ref{Sec211}) is found to substantially decrease the probability of stakeless matches \citep{Gyimesi2024}.

According to Section~\ref{Sec31}, the tournament outcome is generally less uncertain if more games are played. Thus, SKO provides more suspense than other formats, which can explain why many competitions include post-season playoffs \citep{bojke2007impact}. The number of players qualifying for the playoffs also has an impact on suspense. \citet{Olson2014} study the 2014/15 NCAA playoffs that featured only the two strongest teams from the regular season. Suspense is measured at any time period as the expected change in the teams’ probabilities of being champion, summed across ranked teams. A playoff format with more than two teams is found to decrease overall suspense, because the gain in suspense in the playoffs resulting from playing more games does not counter the drop in suspense during the regular season.  

In an incomplete round robin tournament, the number of matches between two teams is also a decision variable. \citet{Lenor2016} and \citet{Kyngas2017} discuss how more popular matches (e.g.\ between local rivals) are selected to occur more frequently in the Australian Football League in order to boost attendance. \citet{Lenor2016} claim they could -- at least as a short-term effect -- increase attendance by 7\% by programming more rivalry games at the expense of the randomly selected matches. 

The number of participating teams is also an important consideration with respect to the attractiveness of the tournament format. In football, the practice seems undecided on this matter: of the 25 European football leagues reported in \cite{GoossensSpieksma2012b} for the 2008/09 season, 8 countries have increased their number of participants by the 2023/24 season, while 5 decreased it.
A structural change happened in the basketball EuroLeague in the 2016/17 season, as the number of participating teams has been reduced from 24 to 16, while the number of games per team has increased. \citet{dimattia2023} find that this reform decreased attendance by about 9 percent per game, predominantly in the first half of the new DRR with 16 teams where the matches are probably perceived as less important. 

\section{Seeding} \label{Sec4}

Seeding is typically used to avoid the top-ranked teams eliminating each other in the earliest rounds. Hence, round robin and Swiss-system tournaments do not require seeding but knockout tournaments, as a stand-alone tournament or as part of a multi-stage tournament, call for seeding. Naturally, the seeding policy may affect our three design criteria.

%Knockout tournaments are usually seeded, that is, the players are ranked based on historical performances, and the brackets are constructed to achieve a given aim.

%A popular way of organizing tournaments with a large number of participants over a limited period of time is to organize it in two stages, where the first stage is organized as a group stage and the second stage as a single knockout tournament. The group stage is commonly organized as a SKO or DKO, where a subset of teams from each group qualify for the next stage. Typically, players are seeded in the second stage according to their performance in the group stage. We speak of a bracket when the allocation of the group position in the knockout tree is fixed beforehand. Figure \ref{Fig3} shows the bracket of the \dots. When the tournament only has one stage, however, the seeding of players typically depends on their historical performance in other tournaments, manifested, for example, in an overall ranking.

\subsection{Efficacy} \label{Sec41}

Seeding can directly enhance the ability of a tournament to select the best players as the winners. However, a misaligned seeding policy might easily favor a weaker player compared to a stronger one. 

\subsubsection{Single knockout tournaments} \label{Sec411}

Perhaps the most popular seeding method is the standard seeding, see Figure~\ref{Fig1a}.
%Section \ref{Sec31} revealed that standard seeding greatly increases the effectivity and efficacy of knockout formats.
For four players, the seeding ((1,4),(2,3)) is generally considered to be the most efficacious option as the probability of winning will be a monotonic function of the rank of players \citep{horen1985comparing, Groh2008}. For a SKO with eight players and $\pi_{ij} = r_i / \left( r_i+r_j \right)$, the standard seeding (((1,8),(4,5)),((2,7),(3,6))) is not the most efficacious: the second strongest player has a higher probability to advance to the finals than the strongest player \citep{prince2013designing}. The reason is that the top player always plays against a player ranked 4 or 5 in the semifinals, while the second best player has a positive probability of playing against the player ranked 6, which increases its chance of reaching the final. Therefore, the seedings (((1,8),(2,7)),((3,6),(4,5))) and (((1,8)(3,6)),((2,7)(4,5))) can be more efficacious than the standard seeding \citep{prince2013designing}. However, \citet{horen1985comparing} unexpectedly find that up to eight different seedings can maximize the probability of the best player winning the tournament. All eight seedings are characterized by player 1 being matched with either player 7 or 8 in the first round, and players 2, 3 and 4 being in the other side of the bracket as 1.

\citet{baumann2010anomalies} observed empirically that the 10th and 11th seeds typically progress further in the playoffs of the NCAA basketball tournament compared to the 8th and 9th seeds. Although the 8th and 9th seeds have a higher probability of winning in the first round, they almost certainly play against the first seed in the second round. In contrast, the 10th and 11th seeds have a more difficult first round but will meet the 2nd or 3rd seed in the next round. Hence, the advantage in the second round seems to outweigh the disadvantage in the first round. This also shows why fairness is sometimes related to efficacy.

%Other seeding methods have also been studied.
Intuitively, random seeding can easily make a stronger player's winning probability smaller than that of a weaker player, since the stronger players might be unequally distributed over both sides of the bracket \citep{israel1981stronger}. This is especially true for an unbalanced KO. However, if the pairwise winning probabilities of players are \emph{strongly stochastic transitive}, namely, $\Pi$ is such that $i < j$ implies $\pi_{ik} > \pi_{jk}$ for all $k$ and $\pi_{kk} = 1/2$, \citet{chen1988stronger} prove that balanced SKOs do not suffer from this anomaly. The condition of strong stochastic transitivity is crucial since the result does not hold for an arbitrary winning probability matrix \citep{Arlegi2020}.  Therefore, \citet{vu2011fair} argue that efficacy can be guaranteed only if the seeding is allowed to vary according to the win probabilities, while \citet{Hwang1982} advocates to use reseeding, i.e.\ seeding (re)done in every round depending on the relative ranks of the players in that round.

%On the other hand, the seeding (13):(24) maximizes the probability of a final among the two top players and has the advantage that it yields the highest expected total effort.
% Krakel (2014) considers only the highest expected total effort, which is not efficacy
%According to \citet{krakel2014optimal}, this depends on the strength distribution of the players, as the seeding (12):(34) maximizes total effort if players 1 and 2 are very strong and players 3 and 4 are very weak. 

\citet{hennessy2016bayesian} also investigate the optimal seeding that maximizes the probability of the best player winning the SKO but only partial information is assumed to be known about the strengths of the players. For 16 players, the optimal bracket places the seven weakest players on the side of player 1. Naturally, a different utility function can lead to another optimal seeding. For example, the seeding that maximizes the probability of the best two players meeting in the final is ((((1,16),(13,14)),((6,7),(8,9))),(((2,15),(11,12)),((3,10),(4,5)))). Since it is computationally prohibitive to obtain the exact values for 16 players, Monte Carlo simulation is used together with simulated annealing to estimate the utility functions. \citet{hennessy2016bayesian} show that the most efficacious design depends on the subset of players that must be ranked according to their true strength.

\citet{karpov2016new} proposes the so-called equal gap seeding, which is (((1,5)(3,7)),((2,6)(4,8))) for eight players. Here, the gap between the strength of players is balanced over all matches. The equal gap seeding can be characterized by a set of reasonable axioms and is the unique seeding that maximizes the probability that the strongest participant is the winner, the strongest two participants are the finalists, etc.\ under the assumptions of the paper.

\citet{Karpov2018} formulates desirable properties for seeding in generalized knockout tournaments where one match involves three or four players. Both the standard and the equal gap seeding are extended to this case and justified by a set of axioms.

\begin{table}[t!]
    \centering
\caption{Studies on the efficacy of seeding policies}
\label{Table2}
\begin{subtable}{\textwidth}
    \centering
    \caption{Papers focusing on efficacy}
    \label{Table2a}
\rowcolors{1}{}{gray!20}
\begin{tabularx}{\textwidth}{lcCc} \toprule
    Study & Best & Worst & $n$ \\ \bottomrule
    \citet{Hwang1982} & Reseeding & Standard & arbitrary\\
    \citet{mcgarry1997efficacy} & (((1,8),(5,4)),((2,7),(6,3))) & (((1,8),(6,7)),((2,3),(4,5))) & 8 \\
    \citet{Groh2008} & ((1,4),(2,3)) & Non-seeded & 4  \\
    \citet{scarf2011numerical} & Reseeding & Random & 32  \\
    \citet{karpov2016new} & Equal gap & Standard & arbitrary \\ \toprule
    %\citet{cohen2023optimal} & 4 & (14):(23) & (12):(34)\\ \toprule
\end{tabularx}
\end{subtable}

\vspace{0.5cm}
\begin{subtable}{\textwidth}
    \centering
    \caption{Papers focusing on effectivity}
    \label{Table2b}
\rowcolors{1}{}{gray!20}
\begin{tabularx}{\textwidth}{lccC} \toprule
    Study & Best & Worst & $n$  \\ \bottomrule
        \citet{searls1963probability} & (((1,8),(6,7)),((2,3),(4,5))) & (((1,3),(2,8)),((4,7),(5,6))) & 8\\
        \citet{horen1985comparing} & (((1,7/8), (5/6, 6,7)),((2, 3/4), (x,x))) & Non-seeded & 4, 8 \\
        \citet{hennessy2016bayesian} & \Centerstack{((((1,16),(14,15)),((10,11),(12,13))), \\ (((2,3)(4,5)),((6,7)(8,9))))} & Random & 16\\ \toprule
\end{tabularx}
\end{subtable}

\end{table}

A structured overview of the papers analyzing seeding rules with respect to efficacy are provided in Table~\ref{Table2}.
%\colorbox{orange}{Value?}.

\subsubsection{Multi-stage tournaments} \label{Sec412}

In the playoffs for the 2020 UEFA European Football Championship, 16 teams were divided into four \emph{parallel} SKOs (paths) with four teams each based on the results of the 2018/19 UEFA Nations League. \citet{Csato2020e} defines a partial ranking on the set of these SKOs.
The rules intend to favor UEFA Nations League group winners, creating appropriate incentives in this tournament. Still, they do not exclude a group winner playing in a more difficult playoff path than a non-group winner of the same league, which is not efficacious. The problem can be solved in an illuminating way: if a group winner envies the position of a non-group winner of its league, then its envy is justified and the group winner can request an exchange of the two teams.
The play-offs in the European section of the 2022 FIFA World Cup qualification and the qualification for the 2024 UEFA European Championship suffer from the same theoretical shortcoming.

Seeding systems used in practice have also been extensively investigated with respect to efficacy. \citet{DagaevRudyak2019} and \citet{CoronaForrestTenaWiper2019} both analyze a reform in the seeding of the UEFA Champions League group stage by Monte Carlo simulation. The changed policy, in use from season 2015/16 to 2023/24, placed the champions of the strongest national associations in the first pot rather than the highest-ranked clubs. While this has only marginally worsened efficacy, the impact on individual clubs has been more powerful, sometimes modifying the probability of qualification for the Round of 16 by 10-15 percentage points.

On the other hand, the empirical study of \citet{engist2021effect} has arrived at a remarkable conclusion based on the results of the UEFA Champions League and UEFA Europa League until 2019/20. In particular, even though a better seed implies weaker opponents on average in the group stage, it does not lead to a higher probability of qualifying for the knockout stage. In other words, seeding itself does not affect efficiency and efficacy. Consequently, it would be especially welcome to carry out similar regression discontinuity studies for other sports and tournaments.

Under the standard seeding, players should theoretically exert full effort in the group stage since this will lead to easier opponents at the beginning of the knockout stage. Nonetheless, the conventional bracket suffers from several weaknesses: the role of luck seems to be excessive and players have an incentive to deliberately lose in some situations \citep{HallLiu2024}. Therefore, it has been suggested to allow higher-ranked players to choose their opponent sequentially in each round \citep{Guyon2022a}. This mechanism can increase the probability that the strongest contestant wins and the two strongest contestants play the final, as well as reduce shirking (i.e.\ deliberately losing a game in order to obtain an easier path through the tournament) \citep{HallLiu2024}.

A variant of the ``choose your opponent'' policy is used in skiing sprint elimination tournaments, however, the athletes seem to make suboptimal decisions \citep{LunanderKarlsson2023}.
The International Central European Hockey League also uses this rule to some extent. In particular, the teams ranked seventh to tenth after the regular (4RR) season play pre-playoffs where the seventh-placed team can choose whether to play the ninth or tenth-placed team. Analogously, the top three teams sequentially select their opponent in the quarterfinals, but the fourth-ranked team is not available to choose.
Finally, the host Egypt had the right to assign themselves to a group of their choice in the 2021 World Men’s Handball Championship \citep{IHF2020}.

\subsection{Fairness} \label{Sec42}

Equal treatment of equals is closely related to efficacy in knockout tournaments according to Section~\ref{Sec41}, so this topic will only shortly be discussed here. On the other hand, it poses a great challenge for certain hybrid tournaments.

\subsubsection{Single knockout tournaments} \label{Sec421}

While a balanced SKO may violate the efficacy principle of stronger players having a higher probability of winning the tournament, the seeding ((1,4)(2,3)) guarantees that equally strong players are treated equally in knockout tournaments with four players \citep{Arlegi2020}.

\citet{Schwenk2000} defines a SKO fair if the probability of winning the tournament is proportional to the inherent strength for each player. However, the author adds that this condition should not be guaranteed by receiving a favorable schedule, a property called \emph{favoritism minimized}. In order to satisfy this requirement, the so-called \emph{cohort randomized seeding} is suggested that involves random seeding and contains $k$ cohorts for $2^{k}$ players. The first cohort consists of the two ``best'' players, i.e.\ $C_{1} = \{ 1, 2 \}$. Then, $C_i = \{2^{i-1}+1, 2^{i} \}$ for all $2 \leq i \leq k$. The seeding of the SKO is designated by the numbers $1,\dots,k$. A random draw determines the assignment of players to the cohorts. This method improves efficacy and still adheres to the desirable properties of knockout tournaments.

\subsubsection{Multi-stage tournaments} \label{Sec422}

To create appropriate incentives to perform in the group stage, the subsequent knockout stage typically matches group winners with teams that performed worse in the group stage. However, this natural structure can be achieved only if the number of groups is a power of two, which might be prohibited by time constraints.
A prominent example is given by a tournament with 24 teams, with a group stage played in six groups of four teams. In football, three FIFA World Cups (1986, 1990, 1994), the UEFA European Championship since 2016, as well as the 2015 and 2019 FIFA Women's World Cups follow this format. The six group winners and the six runners-up qualify for the Round of 16, together with the four best third-placed teams. Thus, some group winners should be matched with a runner-up, and some with a third-placed team in the Round 16.

\begin{figure}[t!]
\centering
\scalebox{0.85}{
\begin{tikzpicture}[
  level distance=4cm,every node/.style={minimum width=2cm,inner sep=0pt},
  edge from parent/.style={ultra thick,draw},
  level 1/.style={sibling distance=6cm},
  level 2/.style={sibling distance=3cm},
  level 3/.style={sibling distance=1.5cm},
  legend/.style={inner sep=3pt}
]
\node (1) {\Pair{FI}{$\mathcal{W}$/SF1}{$\mathcal{W}$/SF2}}
[edge from parent fork left,grow=left]
child {node (2) {\Pair{SF1}{$\mathcal{W}$/QF1}{$\mathcal{W}$/QF2}}
child {node (3) {\Pair{QF1}{$\mathcal{W}$/R1}{$\mathcal{W}$/R2}}
child {node (4) {\Pair{R1}{A2}{C2}}}
child {node (4) {\Pair{R2}{D1}{B/E/F3}}}
} % (3)
child {node {\Pair{QF2}{$\mathcal{W}$/R3}{$\mathcal{W}$/R4}}
child {node {\Pair{R3}{B1}{A/C/D3}}}
child {node {\Pair{R4}{F1}{E2}}
} % (3)
} % (2)
} % (1)
child {node {\Pair{SF2}{$\mathcal{W}$/QF3}{$\mathcal{W}$/QF4}}
child {node {\Pair{QF3}{$\mathcal{W}$/R5}{$\mathcal{W}$/R6}}
child {node {\Pair{R5}{C1}{A/B/F3}}}
child {node {\Pair{R6}{E1}{D2}}}
} % (3)
child {node {\Pair{QF4}{$\mathcal{W}$/R7}{$\mathcal{W}$/R8}}
child {node {\Pair{R7}{A1}{C/D/E3}}}
child {node {\Pair{R8}{B2}{F2}}
} % (3)
} % (2)
};
\node[legend] at ([yshift=2.5cm]4) (R16) {\textbf{Round of 16}};
\node[legend] at (3|-R16) {\textbf{Quarterfinals}};
\node[legend] at (2|-R16) {\textbf{Semifinals}};
\node[legend] at (1|-R16) {\textbf{Final}};
\end{tikzpicture}
}

\caption[The traditional knockout bracket of football tournaments with 24 teams]{The traditional knockout bracket of football tournaments with 24 teams \citep[Figure~5.1]{Csato2021a} \vspace{0.25cm} \\
\footnotesize{Teams are on the right hand side, for instance, A2 is the runner-up of Group A, while B/E/F3 is the third-placed team from Group B, or Group E, or Group F, depending on the allocation rule of the third-placed teams. \\
$\mathcal{W}$ indicates the winner of the previous round.}}
\label{Fig3}
\end{figure}

The traditional design, used until 2016 and depicted in Figure~\ref{Fig3} suffers from severe shortcomings \citep{Guyon2018a, Csato2021a}. Indeed, the groups are not treated equally; the winner and runner-up of Group A (E) enjoy a much easier (tougher) road to the final compared to the other groups, which violates equal treatment of all players (especially if the host is automatically assigned to Group A). Furthermore, the policy used to assign the four third-placed teams to the matches R1--R8 in the Round of 16 is unfair, as the probability of ending up in a harsh quarter (i.e.\ where one meets a group winner in the quarterfinals) is highly unbalanced over the groups. 
Eventually, the 2019 FIFA Women’s World Cup has exchanged some matches in the knockout bracket \citep{Csato2021a} to mitigate these issues, and UEFA has implemented the recommendation of \citet{Guyon2018a} in the 2020 and 2024 UEFA European Championships.

%To summarise, the design of the UEFA European Championships has three advantages over the design of the 2019 FIFA Women's World Cup \citep[Chapter~5]{Csato2021a}:
%\begin{itemize}
%\item 
%The allocation rule for the third-placed teams is not biased;
%\item 
%A repeated matchup (before the final) can occur with a lower probability;
%\item 
%This repeated matchup may happen in the quarterfinals between the winner and the third-placed team of a group, not in the semifinals between the two top teams of a group.
%\end{itemize}

However, this issue remains relevant due to the recent decision of FIFA \citep{FIFA2023} to play the 2026 FIFA World Cup with 48 teams, in 12 groups of 4, essentially ``doubling'' the previous tournaments with 24 teams. Thus, similar to UEFA, FIFA can benefit from the proposal of \citet{Guyon2018a} for future World Cups.

\subsection{Attractiveness} \label{Sec43}

There is an obvious trade-off between efficacy and suspense in knockout and multi-stage tournaments: suspense is heavily reduced in the earlier rounds by deliberately matching the strongest and the weakest teams \citep{ely2015suspense}.
\citet{karpov2016new} compares the equal gap seeding (Section~\ref{Sec411}) with the standard seeding and shows that the former results in a higher competitive intensity.
\citet{Dagaev2018} consider seedings in a SKO with eight players concerning the quality and competitive intensity of the matches. Perhaps unsurprisingly, the seeding (((1,2),(3,4)),((5,6),(7,8))) is optimal for competitive intensity, and the seeding (((1,8),(2,7)),((3,6),(4,5))) maximizes the overall quality of the tournament.
%\citet{Csato2020b} considers the expected quality and competitive balance of alternative tournament formats for the EHF Men's Handball Champions League.

\citet{placek2023impossibility} interprets competitive intensity similarly to \citet{Dagaev2018}. Fairness is defined by minimizing the sum of the ranks of the winners in each match since, in a fair tournament, any player performs at least as well as the lower-ranked players. An impossibility result is derived: under some reasonable conditions, a tournament cannot optimize both competitive intensity and fairness.

\section{Draws} \label{Sec5}

A draw refers to the allocation of teams to groups or knockout brackets, thus, it certainly affects efficacy (Section~\ref{Sec51}) and attractiveness (Section~\ref{Sec53}).
Nonetheless, tournament organizers usually want to retain some randomness in this procedure in order to maximize public interest and carry out a draw as a live media show, which may have a price in fairness as presented in Section~\ref{Sec52}.

\subsection{Efficacy} \label{Sec51}

%As already noted in section in section \ref{Sec422}, when a subset of the third best teams is allowed to qualify, the group draw will have a direct impact on the efficacy of the tournament. This because not all group qualifiers will face the same opponent strength in the subsequent knockout stage. As a result, to maximize efficacy, the best teams should be placed in the groups whereof the group winners have the easiest route to the final. Furthermore, note that balancing the strength of the groups as much as possible will also have a positive effect on the efficacy. Surely, if one group contains only strong and one group only weak teams, at least one team will have a higher chance to qualify than a stronger team. Therefore, although not explicitly mentioned in the literature, we argue that resolving the next two fairness issues also enhance the efficacy of the tournament.

The FIFA World Cup draw has received a lot of criticism in the last decade for creating imbalanced groups, mainly due to the misaligned policy of guaranteeing the geographic diversity of the groups.
\citet{lapre2022quantifying} empirically assess competitive balance between the groups in the FIFA Women’s World Cups from 1991 to 2019, while \citet{LaprePalazzolo2023} repeat this analysis for the FIFA World Cups organized between 1954 and 2022. The groups of both tournaments are shown to exhibit substantial competitive imbalance.

Several solutions have been proposed in the literature to create groups with approximately equal strength subject to geographic constraints. \citet{Guyon2015a} suggests improving the draw procedure with an S-curve-type algorithm. \citet{laliena2019fair} develop another algorithm that outperforms the S-curve-type draw system of \citet{Guyon2015a}. Both proposals involve a heuristic extraction of teams from pots. \citet{laliena2019fair} also provide a procedure for perfectly balanced groups, where the strength of each group coincides, by finding all equal-sum partitions first and drawing one of them randomly. %Since the groups are typically evaluated based on their three strongest teams, only the sum of the three strongest teams' ranks is required to be equal. %Table~\ref{Table8} gives an example of one such equal-sum partition, where the ranks of the first three teams sum up to 38. 

The traditional solution to create $k$ balanced groups is assigning the teams to $k$ pots based on their strength such that teams of similar strength are in the same pot. 
Then, each group will contain a randomly drawn team from each pot. FIFA has introduced this procedure from the 2018 FIFA World Cup \citep{Guyon2018d}.
However, a necessary precondition of balancedness is the appropriate classification of the teams into the pots. This cannot be guaranteed if the underlying ranking -- such as the previous FIFA World Ranking -- is inaccurate, which will be shortly discussed in Section~\ref{Sec71}.

Sometimes the identity of the participants is not known at the time of the draw. For example, the three winners of the playoffs have automatically been assigned to the weakest Pot 4 in the 2022 FIFA World Cup draw. This seeding policy has not balanced the strengths of the groups to the extent possible: a better alternative would have been to assign the placeholders according to the highest-ranked potential winner, similar to the rule used in the UEFA Champions League qualification \citep{Csato2023d}.

Entry to a tournament may be determined by geography. The most prominent example is the FIFA World Cup, where the six continental confederations of FIFA are responsible for their qualifying tournaments. However, the current system is based neither on ensuring the participation of the best teams in the world, nor does it fairly allocate slots by the number of teams per confederation or any other reasonable metric \citep{stone2016unfair}. Hence, exploiting the opportunity offered by the increase in the number of teams in the FIFA World Cup from 2026, \citet{KrumerMoreno-Ternero2023} explore the allocation of additional slots among continental confederations with the standard tools of fair allocation.
\citet{CsatoKissSzadoczki2024} extend the methodology of the FIFA World Ranking to compare the performance of FIFA confederations for a merit-based allocation of qualifying slots among them.
\citet{Csato2023c} investigates the fairness of the 2018 FIFA World Cup qualification to verify that the qualifying probabilities of national teams differ substantially between continents and do not reflect the Elo ratings. For example, the move of Australia from the Oceanian to the Asian zone has increased its probability of playing in the 2018 FIFA World Cup by about 65\%. A partial remedy is proposed by fixing the draw of the inter-continental playoffs.
% Thus, the FIFA World Cup qualification does certainly not satisfy equal treatment of equals.

Balancedness can also be achieved through optimization. \citet{cea2020analytics} provide a mixed integer linear program to balance the strength of the groups and satisfy all geographical constraints. The model is applied for the 2014 FIFA World Cup, and is shown to create groups with more balanced relative strengths compared to the original draw. \citet{melkonian2021fair} gives integer programming formulations for organizing recreational doubles tennis competitions such that for every player, the average ranking of partners in all matches is as close as possible to the average ranking of the opponents in all matches. Nevertheless, an optimization model can have multiple optimal solutions, although generating all of them may be computationally challenging. In this case, rather than simply letting the solver pick one, one should fairly select an optimal solution, which may again involve a lottery (see e.g.\ \citet{Demeulemeester2025}).
For fairness and transparency reasons, however, most organizers stick to a draw.

\subsection{Fairness} \label{Sec52}

Dividing teams from $k$ pots into $k$ groups such that each group contains exactly one team from each pot can be done easily and spectacularly, by drawing the teams sequentially in a random order. However, the problem becomes more complicated if some assignments are invalid; for example, there are geographic restrictions in the FIBA Basketball World Cup draw to maximize the number of inter-continental matches.
In such a constrained draw, fairness has a straightforward definition: each feasible allocation should be equally likely, namely, the draw should be uniformly distributed. Otherwise, teams playing against stronger opponents in expected terms may feel cheated.
%Moreover, the main objective of the draw is to balance the groups as much as possible: a player should not be (dis)advantaged by being placed in a disproportionately (strong) weak group. Hence, the distribution of the strength of teams should be as even as possible for each group. It is clear that the draw is mainly concerned with the fairness issue of treating players equally, but we first make a connection to efficacy.

This property has been violated in the 1990 \citep{Jones1990}, 2006 \citep{RathgeberRathgeber2007}, and 2014 \citep{Guyon2015a} FIFA World Cup draws because the pre-assignment of certain teams to certain groups was combined with the so-called group-skipping policy (i.e.\ if a draw restriction prevents the team drawn to be assigned to the first empty slot in a group, the next group is considered, and so on, until a group is found where the team can be assigned to) to ensure geographical diversity.

A similar problem emerged in the UEFA Champions League, where the Round of 16 draw was implemented with the following constraints between the 2003/04 and 2023/24 seasons:
(a) the eight group-winners have to play against the eight runners-up;
(b) teams from the same group cannot play against each other;
(c) teams from the same national association cannot play against each other.
UEFA used a computer-assisted but transparent draw procedure to enforce the imposed constraints. A probability calculator for the 2023/24 season is available at \url{https://julienguyon.github.io/UEFA-draws/}, and the underlying mathematics -- essentially, Hall's marriage theorem -- is explained by e.g.\ \citet{WallaceHaigh2013, GuyonMeunier2023}.
The mechanism used by UEFA is remarkably close to achieving the fairest possible outcome \citep{BoczonWilson2023}. Nonetheless, while the distortions are small in terms of differences in pairwise match probabilities, in almost every season there exist teams with an expected loss of over 10 thousand euros in prize money due to the bias of the draw procedure \citep{KlossnerBecker2013}.

The UEFA draw procedure has inspired \citet{Guyon2014a} to suggest three sequential mechanisms for the FIFA World Cup draw. A variant of one recommended mechanism has been adopted for both the 2018 \citep{Guyon2018d} and 2022 \citep{Csato2023d} FIFA World Cup draws, and has also been used in the 2019 and 2023 FIBA Basketball World Cups. However, it is also non-uniformly distributed, and the biases are non-negligible, sometimes exceeding one percentage point in the probability of being the group winner or the runner-up \citep{Csato2024g}. Therefore, \citet{RobertsRosenthal2024} propose some fair (uniformly distributed) draw methods that use balls and bowls to remain suitable for a televised event. These solutions can be tried for both the 2018 and 2022 FIFA World Cups at \url{http://probability.ca/fdraw/}.

%The two field-proven draw procedures used in the FIFA World Cup and UEFA Champions League do not coincide. In addition, with $k$ pots, each of them has $k!$ variants since the order of the pots can be arbitrary. \citet{Csato2023m} compares these draw procedures for reasonable subsets of balanced bipartite graphs up to 16 nodes. The design of the UEFA Champions League Round of 16 draw turns out to be the best among the four available options.

A similar draw mechanism is used for the draw of the league phase of the UEFA Champions League and UEFA Europa League, introduced in the 2024/25 season, which are organized as an incomplete round robin tournament with the following constraints:
(a) each of the 36 teams plays eight matches, one home and one away match against two different teams from each of the four pots;
(b) teams from the same national association cannot play against each other;
(c) no team is allowed to play against more than two teams from the same national association; and
(d) the matches should be played over eight matchdays.
A simulator of the draw procedure is available at \url{https://julienguyon.pythonanywhere.com/uefa_app/} and the problem is discussed by \citet{GuyonMeunierBenSalemBuchholtzerTanre2024}.

Draw constraints may threaten fairness even if the draw is uniformly distributed. For instance, consider a SKO with three strong players 1, 2, 3, and one weak player 4. Assume that players 1 and 2 cannot be drawn against each other in the semi-final due to a draw constraint. Then, players 1 and 2 will meet the weak player 4 with a probability of 50\%, but this is impossible for player 3, which violates the principle that equally strong players are treated equally. \citet{Csato2022d} attempts to uncover the effects of draw restrictions on knockout tournaments. Nonetheless, both the quest for fairer draw procedures and the evaluation of existing mechanisms are far from completed.

\subsection{Attractiveness} \label{Sec53}

A disadvantage of striving for groups of equal strength is that it may result in several matches with low competitive intensity and low uncertainty of outcome. Therefore, some tournaments have been intentionally designed with imbalanced groups. The 2024 European Water Polo Championships divided the teams into two divisions based on their past performance. Division 1 (2) contained the first (second) eight teams that played in two groups of four teams each. The group winners and runners-up in Division 1 advanced to the quarterfinals, while the third- and fourth-placed teams had to play against the group winners and runners-up in Division 2 in a playoff round to qualify for the quarterfinals. A similar design was used in the EHF (men's handball) Champions League between the 2015/16 and 2019/20 seasons.
According to the results of \citet{Csato2020b}, this format can lead to more exciting matches without sacrificing fairness.
The UEFA Nations League, introduced in 2018, follows a similar idea, too: the national teams are grouped into four leagues of different strengths with a promotion and relegation system \citep{Csato2022a, ScellesFrancoisValenti2024}.

Balanced groups reduce the competitive intensity of the group stage in order to increase the quality and competitive intensity in the next stage. As far as we know, it has never been considered whether this trade-off is the best solution with respect to the overall quality and competitive intensity of the tournament. According to \citet{Olson2014}, although organizing the playoffs with more players can increase suspense in the playoffs, the total suspense over the whole tournament actually decreases. It is nevertheless desirable if the most attractive matches are played near the end, culminating in an exciting final match.

Draw constraints can also be used to increase attractiveness. For example, FIFA maximizes the number of inter-continental games during the group stage of the FIFA World Cup to make it a unique tournament with matches never seen in continental championships.
\citet{Csato2022a} shows that restrictions in the group draw can decrease the probability of matches that are potentially vulnerable to manipulation, and the suggested framework may be used to avoid any kind of matches specified by the organizer.
However, according to Section~\ref{Sec52}, the usual draw mechanisms cannot guarantee uniform distribution in the presence of draw constraints. A thorough analysis of this trade-off poses an interesting direction for future research.

%\subsubsection{Balanced strength groups}
%\citet{melkonian2021fair} gives integer programming formulations for scheduling doubles tennis competitions such that for every player, the average ranking of partners in all matches is as close as possible to the average ranking of the opponents in all matches. 

\section{Scheduling} \label{Sec6}

% \citet{kendall2017sports} has been removed from the sentence below since this paper is already cited and does not deal with scheduling.

The schedule of a tournament designates the order of the matches. Typically, the schedule is subject to many different constraints from various stakeholders that should be satisfied \citep{ribeiro2012scheduling,Rasmussen2008,duran2021sports}. Some constraints may be \emph{soft}: they can be violated but this comes with a penalty, which allows trade-offs between different desiderata.
Developing a schedule for a single round robin tournament without any constraint is easy \citep{Dewerra1980}. However, the problem becomes $\mathcal{NP}$-hard as soon as some costs need to be minimized \citep{Briskorn2010}.
%The corresponding mathematical model usually aims to minimize or maximize an objective function subject to a set of constraints.

\subsection{Efficacy} \label{Sec61}

Compared to other design criteria, the scheduling literature is much less concerned about efficacy. Nevertheless, the schedule may have an impact on the efficacy of the tournament. Several empirical papers investigate whether the order of the matches in two-legged clashes in knockout tournaments matters, but the results are mixed (e.g.\ \citet{Flores2015, Amez2020}). \citet{Jost2024} identifies the conditions under which a team prefers to play first home, first away, or is indifferent, depending on the size of home advantage, the relative strengths of the two teams, and the tie-breaking rule.

In a theoretical work, \citet{Krumer2017} consider a single round robin tournament with three symmetric players and one prize. The player that has no game in round 2 is found to have the highest expected payoff. In the same tournament with two prizes, the sequence of the matches does not affect the winning probabilities if and only if the second prize is valued at half of the first prize \citep{LaicaLauberSahm2021}.
In a single round robin tournament with four players where no games are played simultaneously, there is a \emph{first-mover advantage}: the player who competes first in rounds 1 and 2 enjoys a significant advantage \citep{Krumer2017}.
%because a win for a player in the first match of the first round can alter the motivation of other players to continue the tournament. If player $i$ loses in round 1 against $j$, and $j$ wins against another player $k$ in round 2, $i$ might consider the tournament to be over and stop exerting full effort.
These theoretical predictions are confirmed by historical match outcomes from the FIFA World Cup and Olympic wrestling tournaments \citep{Krumer2017b}.

Even if a round robin tournament is efficacious at the end of the season, there are still concerns whether intermediate rankings reflect the true strength of the opponents (see also Section~\ref{Sec721}). A crucial issue is whether each team has played (approximately) the same number of games. The \emph{games played difference index} \citep{suksompong2016scheduling} is the minimum integer $p$ such that at any moment, the difference in the number of games played between any pair of teams is at most $p$. An integer program minimizing this measure is proposed by \citet{van2020handling}. \citet{GoossensSpieksma2012b} call a schedule \emph{g-ranking-balanced} if the difference between the number of home games played by any two teams up to each round is at most $g$. Ideally, the sequence of the strengths for the opponents of each team is also as balanced as possible over the tournament. Theoretical results about such schedules are given by \citet{briskorn2009combinatorial} and \citet{briskorn2010constructing}. \citet{briskorn2009ip} propose an integer program to distribute the matches of each team against teams of a specific strength group as evenly as possible.

\subsection{Fairness} \label{Sec62}

Fairness is an important concept in sports scheduling, reflected by the objective functions of many studies on constructing sports schedules.
%In their sports scheduling classification,
\citet{van2020robinx} mention travel distance, breaks, and the carry-over effect as possible objectives (besides the more general costs and soft constraints); in this section, we also discuss rest times. Typically, no schedule makes every team happy. Therefore, while most of the literature deals with minimizing the global dissatisfaction with the schedule (i.e.\ criterion efficiency), we particularly focus on how this aversion is balanced over the teams (i.e.\ distribution equity). We refer to \citet{GoossensYiVanBulck2020} for a discussion of the trade-off between both angles in the context of sports scheduling.

\paragraph{Travel distance}

Balanced travel times are particularly important if teams do not face each opponent an equal number of times at home and away (e.g.\ in a $k$RR tournament with an odd $k$), or if a team can make away trips, traveling directly from the venue of one opponent to the venue of the next without returning home -- as in the well-known traveling tournament problem, see e.g.\ \citet{Ribeiro2007, Irnich2010}. 

Surprisingly, the relatively large literature on the traveling tournament problem has focused exclusively on minimizing total travel distance, ignoring the matter of balancing the travel distance over the teams.
\citet{duran2021scheduling} balance the travel distances of the teams in six Argentinean youth football divisions that play a SRR. Their solutions have been effectively implemented since 2018. \citet{osicka2023fair} adopt a cooperative game theoretical framework to allocate the travel distances among the teams according to different notions of fairness. They also study the trade-off between minimizing total travel and maximizing fairness.

\paragraph{Breaks}

%It is well known that teams playing at home benefit from home advantage \citep{pollard2014components, BakerMcHale2018}. In a $k$RR with an odd $k$, teams do not play the same number of home and away games, which can be seen as unfair. Nonetheless, some soccer leagues organized in this format; for example, the Hungarian first division is a 3RR.

%If $k$ is even, the sequence of home and away games determines the fairness of the schedule. This is called the HAP (Home-Away Pattern), the set of HAPs of all teams is called the HAP-set. The sports scheduling literature is especially interested in breaks. 
A \emph{break} occurs if a team plays two consecutive games at home or away, which is considered undesirable. Consecutive home games have a negative impact on stadium attendance \citep{forrest2006new}, and consecutive away games can result in substantial revenue loss \citep{duran2017scheduling}. Breaks are especially deemed unfair at the beginning and the end of the season \citep{GoossensSpieksma2012b}. The group stage of the UEFA Champions League has, for instance, been played without breaks in the first or last rounds for any team.

The definition of a break may depend on other characteristics of the tournament. For example, the qualifications for the FIFA World Cup are usually played in double rounds: the national teams play two matches over a week, followed by a longer period without any matches. Then the focus ought to be only on breaks in these double rounds. In particular, away breaks in double rounds are worth avoiding because they might imply a long period without home games and can mean long travels within just a few days. Therefore, \citet{duran2017scheduling} develop an integer programming model for scheduling the South American Qualifiers for the FIFA World Cup that minimizes the total number of away breaks within double rounds. Their proposal was unanimously approved and has been used since the qualifiers for the 2018 World Cup. \citet{GoossensSpieksma2011} even consider breaks between non-consecutive rounds, motivated by the idea of balancing the number of midweek home games over the teams (see \citet{KrumerLechner2018} and \citet{Goller2020} for empirical evidence that midweek games are different from the usual weekend games).

Several studies attempt to minimize the total number of breaks, both from a practical and theoretical point of view (see e.g.\ \citet{Elf2003, RasmussenTrick2007, van2010constructing, ribeiro2012scheduling, tanaka2016fair, van2020handling, RibeiroUrrutiadeWerra2023b}). Finding an equitable schedule with a minimal number of breaks such that breaks are equally distributed over all teams, has also received some attention. For a SRR with $n$ teams, an equitable schedule where each team has one break exists \citep{Dewerra1980}. \citet{Froncek2005} show that a unique schedule without breaks exists if one extra round is used (and hence each team has one round without a match). \citet{Miyashiro2005} propose a polynomial-time algorithm to find an equitable schedule for a given timetable of a round-robin tournament. 

\paragraph{The carry-over effect}

Player $i$ is said to give a \emph{carry-over effect} to player $j$ if another player $k$ plays against $j$ immediately after playing against $i$ \citep{Russell1980}. It is considered unfair if a player systematically gives a carry-over effect to the same player. Take, for example, the TATA Steel Chess Championship 2020. The winner of the tournament, \emph{Fabiano Caruana}, was ranked second in the world at that time. Remarkably, Caruana played in 10 of the 13 rounds against someone who had played in the previous round against \emph{Magnus Carlsen}, the number one player in the world at that time. According to \citet{TATA2020}, this schedule gave Caruana an unfair advantage, letting him play often against opponents who were ``exhausted'' or ``depressed'' after (losing) their match against Carlsen. On the other hand, \citet{GoossensSpieksma2012a} find no evidence that carry-over has a significant impact on the results of Belgian professional football teams. Nevertheless, substantial effort has been devoted to balancing the carry-over effects as evenly as possible over all players (e.g.\ \citet{Russell1980, anderson1999balancing, trick2000schedule, henz2004global, miyashiro2006minimizing, kidd2010tabu, guedes2011heuristic}).

\paragraph{Rest time differences}

Shorter rest times are related to both increased injury rates \citep{dupont2010effect, bengtsson2013muscle} and higher levels of fatigue. This introduces an element of unfairness if a team with a fully rested squad plays against a team having a short rest time \citep{scoppa2015fatigue}. Indeed, \citet{bowman2022schedule} find that in the NBA, back-to-back games are the single most significant factor affecting team success. An easy way of reducing the differences in rest times between teams is to limit the total number of games per week for each team \citep{knust2010scheduling}. Other approaches are to directly minimize the rest time differences between teams \citep{atan2018minimization, ccavdarouglu2020determining, duran2022mathematical, tuffaha2023round}, or to minimize an aggregated rest time penalty (ARTP) which gives a higher weight to shorter rest times \citep{van2020handling}. Based on the latter, \citet{GoossensYiVanBulck2020} propose an integer program formulation and a bi-criteria evolutionary algorithm to generate a set of so-called equitably-efficient schedules. A schedule is equitably-efficient if no schedule exists with a smaller ARTP for one team, or with a smaller ARTP for a worse-off team at the expense of a better-off team, without worsening the ARTP of the other teams compared to this schedule. They apply and visualize this for a non-professional indoor football league, allowing practitioners to make a well-informed trade-off.\\

Unsurprisingly, there exist unavoidable trade-offs between these fairness criteria. The distance minimization and the break maximization problems are equivalent if the distance between every pair of teams is equal \citep{UrrutiaRibeiro2006}. \citet{craig2009scheduling} focus on the trade-off between minimizing travel distance for teams, avoiding sequences of consecutive home or away games, and ensuring equity in rest times. For a SRR with $n$ teams, the minimal number of breaks is $n-2$, which is achieved (for instance) by the \emph{canonical} schedule. However, this schedule also yields maximally unbalanced carry-over effects for an even number of teams \citep{LambrechtsFickerGoossensSpieksma2018}. \citet{gunnecc2019fair}, \citet{GoossensYiVanBulck2020}, and \citet{CavdarougluAtan2022} discuss the trade-off between minimizing breaks and balancing the carry-over effect. Obviously, there are also trade-offs between fairness and organizational or financial criteria (see e.g.\ \citet{goossens2009scheduling, ribeiro2012scheduling}).

\subsection{Attractiveness} \label{Sec63} % and dynamic scheduling

In the sports scheduling literature, basically two approaches exist to deal with attractiveness. One approach assumes that the set of popular matches is known or given by a broadcaster, possibly inspired by studies like \citet{Wang2018, Krumer2020testing, Jakee2022}. The schedule should then assign as many of these games as possible to commercially interesting timeslots, and avoid any overlap between them. This approach has been reported in several practical applications, from e.g.\ Australian football \citep{Kyngas2017}, football \citep{goossens2009scheduling,ribeiro2012scheduling}, and volleyball \citep{cocchi2018}. \citet{Matsumura2022} estimate the impact of scheduling decisions on TV audiences for the Japanese professional baseball league by multiple regression analysis to develop a schedule that maximizes the expected number of spectators. 
Domestic football competitions occasionally shut down for some days due to periods reserved for international matches played by national teams. According to \citet{Perez2024}, home wins have a lower probability after such an interruption, which leads to greater uncertainty of outcome. Consequently, the set of matches played immediately after ``FIFA reserved dates'' needs to be chosen strategically.

The other, much less attempted approach tries to find a schedule that is as attractive as possible given a prediction of match results, or robust with respect to a subset of possible outcomes. \citet{yi2021dealing} proposes an implementer-adversary approach to dynamically develop round robin schedules that maximize suspense, measured by how late in the season the winner is decided. The algorithm works by generating a set of scenarios (realizations of match outcomes) based on historical data from betting odds. The implementer problem finds a schedule that reveals the winner as late as possible, given the information about match outcomes. Next, the adversary problem finds a new (realistic) scenario under which the generated schedule is the worst possible, and this new scenario is added to the set of scenarios. The algorithm continues until the implementer and adversary solutions converge. 

We see at least two avenues for future research on improving the attractiveness of tournament schedules. The first is to include match importance measures (see Section~\ref{Sec233}) in order to obtain a schedule that maximizes the total importance of matches over the season.
This can be easily done in a SRR with three or four teams; papers focusing on stakeless matches usually recommend an optimal \citep{ChaterArrondelGayantLaslier2021, Guyon2020a} schedule.
Using complete enumeration, \citet{CsatoMolontayPinter2024} have managed to solve this problem in a DRR with four teams and some scheduling constraints. Given the complexity of these measures, such an endeavor will be challenging computationally for larger tournaments.
The second avenue focuses on dynamic scheduling, which means that the schedule is not computed and announced fully at the beginning of the season, but is released gradually during the season. Except for recent applications in the NFL \citep{bouzarth2020dynamically}, in the FIFA World Cup \citep{Stronka2024}, and in e-sports \citep{dong2023dynamic}, dynamic scheduling has not received much attention in the literature. Yet, it may have a great potential to improve the attractiveness of tournaments, as recent information about the strength and ranking of players can be taken into account in order to maximize the importance of the remaining matches.

\section{Ranking methods} \label{Sec7}

%Rankings are often used for seeding and slot allocation. The group composition in the FIFA World Cup depends on the FIFA World Ranking,  In tennis, only the 104 highest-ranked players qualify automatically for the prestigious Grand Slam tournaments.

The theoretical properties of ranking procedures have been widely discussed in the literature.
\citet{gonzalez2014paired} provide a comprehensive axiomatic analysis of ranking methods for general tournaments where the players may play an arbitrary number of matches against each other. Applying maximum likelihood to the (Zermelo--)Bradley--Terry model \citep{Zermelo1929, BradleyTerry1952}, and the generalized row sum, a parametric family of ranking methods introduced by \citet{Chebotarev1994}, turn out to be the most appealing by satisfying all reasonable axioms that are compatible with each other.
%These axioms have inspired \citet{LeivaBertran2022}, too, in the look for fair and internally consistent standards to rank the players in a tournament.

\subsection{Efficacy} \label{Sec71}

While efficacy in the context of tournament formats and seeding rules refers to the ability to rank the players according to their true (but hidden) strength, the efficacy of a ranking method usually refers to its \emph{predictive power}, the ability of a ranking to accurately predict future matches.
%As such, in the ranking literature the term predictive power is more commonly used. Historical rankings can be used to predict subsequent match results.If the past rankings accurately predict future outcomes, the ranking system is said to have a high predictive power, see for example \citet{mchale2007statistical} for an application to the FIFA world rankings.
\citet{mchale2007statistical} test a previous formula of the FIFA World Ranking from this perspective. \citet{lasek2013predictive} compare the predictive power of several methods for ranking national football teams. The Elo rating is found to be a better indicator of success than the old FIFA World Ranking, a result that is reinforced by \citet{GasquezRoyuela2016}. Indeed, FIFA reformed the World Ranking in 2018 \citep{FIFA2018c}. Nonetheless, the predictive power of the current FIFA World Ranking can still be considerably improved by incorporating home-field advantage and introducing a weighting by goal difference \citep{SzczecinskiRoatis2022}. Somewhat analogously, a variant of the Elo rating (Football Club Elo Ratings, \url{http://clubelo.com/}) robustly outperforms the UEFA club coefficient in terms of predictive accuracy for the UEFA Champions League \citep{Csato2024c}.

\begin{table}[t]
\centering
\caption{Papers on the efficacy of ranking methods}
\label{Table5}
%\begin{tabular}{p{.25\textwidth}|p{.15\textwidth}|p{.15\textwidth}|p{.15\textwidth}}
\rowcolors{1}{}{gray!20}
\begin{tabularx}{\textwidth}{lcCC} \toprule
        & Application & Best & Worst \\ \bottomrule
        \citet{govan2008ranking} & \Centerstack{NFL \\ NCAAF \\ NCAAB} & Massey  & Keener \\
        \citet{lasek2013predictive} & National football teams & Elo (\url{eloratings.net}) & Network based \\
        \citet{GasquezRoyuela2016} & National football teams & Elo (\url{eloratings.net}) & \Centerstack{(previous) \\ FIFA World Ranking} \\
        \citet{chartier2011sensitivity} & NFL & Massey, Colley & PageRank \\
        \citet{barrow2013ranking} & \Centerstack{MLB \\ MBA \\ NCAAB \\ NCAAF} & \Centerstack{--- \\ Keener \\ --- \\ Least squares} & \Centerstack{PageRank \\ Win percentage \\ --- \\ ---} \\
        \citet{dabadghao2022predictive} & \Centerstack{NBA \\ NFL \\ NHL} & \Centerstack{(1,$\alpha$) \\ (1,$\alpha$) \\ (1,$\alpha$)} & \Centerstack{Elo \\ Win-loss \\ Win-loss} \\ 
        \citet{Csato2024c} & UEFA Champions League & Elo (\url{clubelo.com}) & UEFA club coefficient \\ \toprule
\end{tabularx}
        %\end{tabular}
\end{table}

In round robin tournaments, the points scored in individual matches are usually summed for each team, and the teams are ordered based on the total number of points obtained (Section~\ref{Sec221}). \citet{dabadghao2022predictive} call this the win-loss method, which is compared to other, more sophisticated ranking methods for the NBA, NHL, and NFL. Although the win-loss method is used officially in these leagues, it is the worst performing method for the NFL and NHL, and the second worst for the NBA. An extension of the Markov method, which is a pairwise comparison ranking method that uses Markov chains to rate and rank the competitors, turns out to be the most efficacious in each of the three cases.
\citet{govan2008ranking} and \citet{barrow2013ranking} also compare ranking methods in North American sports competitions.
\citet{TiwisinaKulpmann2019} develop a statistical model that assumes only transitivity to describe the outcomes of sports matches, and applies it to evaluate ranking methods. Scoring systems (3,1,0) and (2,1,0) are found to be close to each other and competitive with the Elo rating in the German football league and NFL. 
%Since a discussion about the mechanics of each of the ranking method would lead us to far, we do not include them here but instead refer to these papers. 
Table~\ref{Table5} presents a concise overview of some of these results.

%The construction of a ranking for incomplete round robin tournaments is another popular issue that that has been partially discussed in Section~\ref{Sec22}.

Standard scoring rules perform badly in settings where players play a different number of matches and some pairs of players do not play against each other, which is the case, for instance, in tennis. \citet{radicchi2011best} shows that a network-based ranking method has better predictive power than other well-established methods. \citet{BozokiCsatoTemesi2016} and \citet{TemesiSzadoczkiBozoki2024} use the eigenvector and least squares methods to derive all-time rankings for men and women tennis players, respectively. \citet{ChaoKouLiPeng2018} use a similar methodology to rank players in Go.

%The fact that more sophisticated methods are better predictors should not come as a surprise: the more information about the quality of a win that is taken into account, the more efficacious the ranking method should intuitively be.

\subsection{Fairness} \label{Sec72}

\citet{vaziri2018properties} introduce three criteria for the fairness of a ranking:
(a) the quality of a victory or defeat is taken into account (opponent strength);
(b) a team always has a clear incentive to win a match to improve its rank (incentive compatibility);
(c) a factor should not be considered if the teams have no direct control over it (independence of the schedule).
Opponent strength is often not taken into account in round robin tournaments, but it can be important in other tournaments.
Incentive compatibility, which is basically equivalent to the monotonicity properties of \citet{gonzalez2014paired}, is not discussed in this survey (see Section~\ref{Sec1}).
One might think that all reasonable ranking methods are independent of the schedule, but this is not the case. In particular, the Elo rating depends on the sequence of matches: if two teams play two games against each other such that one is won by both teams, then winning the second game leads to a higher rating \citep{CsatoKissSzadoczki2024}.

A possible interpretation of fairness is \emph{self-consistency}, which requires assigning the same rank for players with equivalent results, while a player who performs better than another should be ranked strictly higher. This fundamental axiom is however incompatible with two other reasonable properties on the set of general tournaments where the players are allowed to play any number of matches against each other.
The first is \emph{independence of irrelevant matches} \citep{Csato2019d}: the relative ranking of two players should be independent of the outcome of any match played by other players.
The second is \emph{order preservation} \citep{Csato2019e}: if player $i$ is not worse than player $j$ in two tournaments where all players have played the same number of rounds, then their relative ranking should remain the same even when the two tournaments are aggregated.

However, these results do not imply that the axioms cannot be met simultaneously on a restricted set of tournaments. In particular, the three properties above are satisfied in a round robin tournament if the players are ranked according to the number of points scored, provided that a tie is worth half of a win. Note that both the maximum likelihood and the generalized row sum ranking methods are score consistent, they coincide with the ranking based on scores on the set of round robin tournaments \citep{gonzalez2014paired}.

\subsubsection{Round robin tournaments} \label{Sec721}

Ranking in a round robin tournament usually does not necessitate taking the strength of opponents into account because any two players play against the same set of opponents (themselves disregarded). However, this does not hold in an incomplete tournament.

During the COVID-19 crisis, almost all sports competitions were suddenly shut down, and some of them never finished, accidentally turning them into incomplete round robin tournaments.
Obviously, summing all the points obtained over the unfinished season does not result in a fair ranking, since the strength of opponents can substantially differ between the teams.

Thus, several ranking methods that take the strength of opponents into account have been proposed in order to construct a fair ranking. In particular, the method of \citet{gorgi2023estimation} measures the performance of the teams in the previous matches of the season and predicts the remaining non-played matches through a paired-comparison model.
\citet{van2023probabilistic} develop a tool to calculate the probabilities of possible final rankings.
\citet{Csato2021d} identifies a set of desired axioms for ranking in these incomplete tournaments and verifies that the generalized row sum satisfies all of them.

\subsubsection{Knockout tournaments} \label{Sec722}

\citet{Bahamonde-BirkeBahamonde-Birke2023} find that in home-away clashes between two teams, the advantage that the team closing the series at home enjoys can be countered to some extent by valuing away goals over home goals in the event of a tie. However, UEFA decided to abolish this away goals rule in its competitions starting from the 2021/22 season. Note that this issue is also related to scheduling in knockout tournaments, see Section~\ref{Sec61}.

Among all topics of tournament design, penalty shootouts have received probably the most serious attention in the economic literature; highly influential empirical works are \citet{ApesteguiaPalacios-Huerta2010}, \citet{KocherLenzSutter2012}, \citet{Palacios-Huerta2014}, and \citet{KassisSchmidtSchreyerSutter2021}. There are also several suggestions to improve the fairness of penalty shootouts as a tie-breaker by changing the standard shooting order, which are summarized in \citet{CsatoPetroczy2022a}.

\subsubsection{Swiss system tournaments} \label{Sec723}

In a traditional Swiss system tournament, the players are paired using a set of complex rules, which basically aim to ensure that the opponents have a similar running score (see Section~\ref{Sec323}).
Take two players $i$ and $j$ who finish with the same score such that $i$ has scored most of its points at the beginning of the tournament, while $j$ has scored most of its points at the end of the tournament. Consequently, the opponents of $i$ have probably \emph{higher} scores than $i$ (it is unlikely that all of them show the same decline in performance as $i$) and the opponents of $j$ have probably \emph{lower} scores than $j$ (it is unlikely that all of them show the same improvement in performance as $j$).

As we have seen in Section~\ref{Sec221}, the strength of opponents is considered by some tie-breaking rules; for instance, the Buchholz score adds the scores of all opponents.
However, taking the strength of opponents into account only in breaking ties is obviously insufficient. In the situation above, player $j$ might easily score marginally more points than player $i$ by playing against much weaker opponents.
\citet[Chapter~1.2]{Csato2021a} presents an example from the 2011 European Team Chess Championship Open tournament: Serbia scored 10 and France scored 9 points but their Buchholz scores were 152 and 179, respectively. Therefore, ranking Serbia above France can be easily debated. \citet{Csato2013a} and \citet{Csato2017a} have proposed novel methods, however, the optimal trade-off between the number of points scored and the strength of the schedule remains unknown.

\subsection{Attractiveness} \label{Sec73}

It is probably harmful to attendance in any tournament if the final position of a team is already secured, independently of the outcomes of the remaining games. \citet{Csato2023a} uncovers that tie-breaking rules have a non-negligible role in this respect: the chance of a fixed position in the group ranking can be reduced by at least two and usually five percentage points in the last round of a DRR with four teams if goal difference and number of goals scored are the primary tie-breaking criteria instead of head-to-head results.

\citet{garcia2022does} show that interest in Formula One decreases once the driver’s champion is known. Although a natural solution would be to decrease the relative value of winning a race, this would increase the probability that a driver could win the competition without winning any races -- as happened in the 1999 Motorcycle Grand Prix 125cc. \citet{Csato2023b} evaluates point scoring systems for Formula One from this perspective, and finds that the current point scoring system provides a reasonable compromise between these two perils. Furthermore, it is competitive with the geometric scoring rules proposed by \citet{KondratevIanovskiNesterov2023} based on axiomatic arguments.

Until the 1998/99 season, the NHL gave 2 points for a win, 1 point for a draw, and 0 points for a loss. However, starting with the 1999/2000 season, the league awarded one point for a loss in overtime in order to decrease the frequency of overtime games ending in a tie. The new scoring system has achieved the outcome intended by league officials but the ratio of overtime games has also risen as the total reward has increased for games ending in overtime \citep{Abrevaya2004}.

Almost all round robin football leagues award a win by 3 points, a draw by 1 point, and a loss by 0 points. This method was introduced first in England in the 1981/82 season, and adopted by other tournaments mainly in the 1990s.
\citet{Moschini2010} develops a game-theoretic model that reflects the crucial strategic elements of a scoring system where the value of a win is a parameter. The empirical results are in line with the model implications: the (3,1,0) system leads to a significant increase in the expected number of goals and a decrease in the ratio of draws. However, according to \citet{GuedesMachado2002}, a higher reward for victory induces the weaker team to play more defensively if the difference between the strengths of the two teams is sufficiently high. The authors also find empirical support for this hypothesis.

There are many more papers on this issue. However, since an overwhelming majority of them have been published in (sports) economics journals, they are not discussed here. The main message is probably that changing ranking rules in order to increase the attractiveness of the tournament may have unintended consequences. Hence, consultation with the scientific community before rule changes would not be a futile exercise.

\section{Conclusions and directions for further research} \label{Sec8}

Our study has aimed to overview the tournament design literature from the perspective of operational research. A basic question of the field is the efficacy (i.e.\ the ability to rank all players according to their true strength) of tournament designs. In the case of knockout tournaments, seeding is usually found to have a significant effect on both measures, although some works challenge this result. Fairness is also an important aspect of tournament design but has no uniform definition. Many papers call a tournament fair if it favors stronger players, while other studies argue that players should be treated equally. The importance, suspense, and quality of matches are also crucial criteria for choosing a tournament design.

Unsurprisingly, round robin tournaments seem to be better at identifying the true ranking of the competitors, while knockout tournaments require fewer matchdays and do not contain unimportant matches. This may explain why popular sports competitions, like the FIFA World Cup, have adopted hybrid designs consisting of a round robin group stage followed by a single knockout tournament. However, the traditional group stage often becomes ``boring'', hence, there are recent attempts to create imbalanced groups or replace groups with an incomplete round robin tournament.

%Operational research mainly contributes to tournament design by optimizing design choices over certain criteria. However, tournament design as a research field has also a clear interface with sports economics. Indeed, in sports economics, the main question is typically what is the impact of certain design choices on the tournament, through theoretical or empirical research. There is a clear synergy between these angles, where sports economics research can bring new criteria and conditions to the table to be taken into account in optimization (or debunk myths that traditionally play a role in tournament design), and validate the decisions proposed by operational research. Ideally, although this survey has focused on contributions from operational research, it will also be useful for sports economists, and advance collaboration between the two domains.

Based on the previous sections, several issues require further investigation by the operational research community. In particular, we think more research is needed in the following areas:

\paragraph{Optimizing the number of contestants}
What is the optimal number of contestants for a tournament in terms of efficacy and attractiveness? Changes in the number of contestants, increases as well as decreases, have regularly occurred during the recent decades in practice; however, these decisions seem to have rarely been based on scientific research.

\paragraph{Robustness of seeding policies}
Recently, \citet{engist2021effect} have found no empirical evidence that seeding contributes to success in the UEFA club competitions. Since this result is in contrast to the common assumption of the operational research literature, more studies are needed for other tournaments and sports to decide whether it is only a statistical oddity or a more general finding. The robustness of the (optimal) seeding rule also requires further investigation since the true ranking of the players can only be approximated in practice.

\paragraph{Constrained draw}
If draw constraints exist, the currently used random draw procedures do not guarantee that all valid assignments are equally likely. How can we develop draw mechanisms that respect all constraints (e.g.\ geographic, scheduling), are (approximately) fair, but still offer a transparent implementation that is appealing to a TV audience? In particular, the issue is worth studying in the incomplete round robin format, which contains an added complexity: all teams should eventually be ordered in a single ranking, even though the teams have faced different sets of opponents (determined by the draw).

%\paragraph{Tournaments with opponent choice}
%As we have seen in Section~\ref{Sec412}, higher-ranked players can be allowed to choose their opponent sequentially in each round of a knockout tournament \citep{Guyon2020a, HallLiu2024}. Is such a reform able to considerably improve efficacy and fairness? What happens if the players have diverse incentives (e.g.\ sporting vs.\ financial), or suffer from cognitive limitations, pushing them towards suboptimal decisions?

\paragraph{Imbalanced groups versus incomplete round robin tournaments}
It seems that organizers attempt to increase attractiveness and guarantee more matches between the top players by two different formats: deliberately creating imbalanced groups (Section~\ref{Sec53}) or replacing the group stage with a single incomplete round robin tournament (Section~\ref{Sec211}).
It remains an open question to what extent can these formats improve our main design criteria. %Furthermore, they should be formed ``cleverly'' to prevent manipulation, when a team pretends to be weak in order to benefit from the novel designs by playing against weaker opponents.
In particular, the novel incomplete round robin format of UEFA club competitions is foreseen to provide greater fairness, a more dynamic ranking, and a higher level of sporting interest \citep{UEFA2023}. However, these expectations have not been confirmed by a scientific analysis at the moment, except for a recent study focusing on competitive balance \citep{Gyimesi2024}. It would also be interesting to explore how a predetermined set of matches worsens efficacy compared to the Swiss system.

%The trade-off between the ratio of matches that are scheduled dynamically (depending on the results of the previous matches) and other properties

\paragraph{Organising incomplete round robin tournaments}
The incomplete round robin format of UEFA club competitions, introduced in the 2024/25 season, substantially increases the complexity of addressing the theoretical criteria discussed in our survey. The set of matches to be played depends on the seeding rules and is determined by a draw. However, it may lead to an infeasible schedule (if the chromatic index of the regular graph representing the matches exceeds the number of available match days), in other words, the matches-first approach should consider at least some scheduling constraints. On the other hand, the schedule-first approach, when the teams are assigned to the already chosen slots, is too rigid and may threaten the randomness of the draw \citep{GuyonMeunierBenSalemBuchholtzerTanre2024}. Finally, the properties of the ranking method are also affected by the seeding regime and the set of matches to be played. Therefore, the organizer needs to simultaneously take all these aspects into account, and their optimal treatment seems to provide serious challenges for OR in the coming years.

\paragraph{Dynamic scheduling}
%How can the schedule of a tournament be altered/determined \emph{during} the season in order to increase efficacy, fairness, or attractiveness?
What flexibility remains to schedule the last half/quarter of the tournament, once the first part has been played (see \citet{lambers2023flexibility} for some measures)? To what extent can extra information (e.g.\ on which teams are still running for which prizes) be taken into account? What is the impact of (partial) dynamic scheduling on efficacy, fairness, and attractiveness? To reduce practical challenges, the organizers could fix the home game days for each team at the start of the season, and fill in the opponents gradually during the season. How should each team's home dates be set in order to maximize planning flexibility for the later parts of the season?
What is the trade-off between the ratio of matches that are scheduled dynamically and the performance of the tournament with respect to the main design criteria?

\paragraph{Multi-league scheduling}
The extant literature focuses mainly on scheduling a single league or tournament. How to deal with multiple tournaments that are played simultaneously and linked with each other, e.g.\ through shared infrastructure, or players taking part in more than one tournament \citep{davari2020multi, Li2023}? To what extent is scheduling linked with the grouping of teams into leagues (resulting from a draw, promotion/relegation, or optimization over e.g.\ travel distance)?

\paragraph{Optimizing global ranking systems}
How can the ranking methods of sports federations be optimized to predict future performance? Can they be improved by changing their methodology, or taking further factors into account? To what extent can other tournament design choices (format, schedule, draw, etc.) contribute to this accuracy? The issue is important especially because global rankings are widely used for seeding and the allocation of qualifying slots.\\

Tournament design is a research domain that has a strong link with practice. Practitioners and sports governing bodies will hopefully take an interest in this survey, too. In recent years, the results of academic research have frequently been adopted by sports administrators:
\begin{itemize}
\item
FIFA has introduced a better ranking procedure for men's national teams in 2018 \citep{FIFA2018c} (Section~\ref{Sec223});
\item
UEFA has modified the knockout bracket of the 2020 and 2024 UEFA European Championships to balance group advantages according to the recommendation of \citet{Guyon2018a} (Section~\ref{Sec422});
\item
FIFA has reformed the draw of the 2018 and 2022 FIFA World Cups inspired by \citet{Guyon2014a}, see \citet{Guyon2018d} (Section~\ref{Sec52});
\item
The proposal of \citet{duran2017scheduling} for a fairer schedule of the South American Qualifiers for the 2018 FIFA World Cup has been unanimously approved by the participants and is currently being used (Section~\ref{Sec62}).
\end{itemize}

Governing bodies in sports can also learn from each other. Both the FIBA Basketball World Cup and the FIFA World Cup aim to maximize geographic diversity in the group stage since maximizing intercontinental matches is useful both for attendance and ranking purposes \citep{Guyon2015a}. On the other hand, the 2023 World Men's Handball Championship contained a group with two of the three African teams, another group with two of the four South American teams, while the number of European teams varied between one and three in the eight groups.

The example of the 2022 FIFA World Cup draw warns that it is worth consulting with the wider scientific community before rule changes are implemented. On 22 March 2022, FIFA released the details of this draw procedure, which treated the three placeholders of the playoffs unfairly and threatened the balancedness of the groups as discussed in Section~\ref{Sec51}. Even though \citet{Csato2023d} immediately realized the problem, it was clearly impossible to correct the mistake before the draw happened.

Consequently, since governing bodies in sports quite often change the design of their tournaments, usually driven by commercial motives, and some innovative ideas repeatedly emerge in practice, operational research in sports will certainly remain a growing research field for the coming decades.

\section*{Acknowledgements}
\addcontentsline{toc}{section}{Acknowledgements}
\noindent
The research was supported by the National Research, Development and Innovation Office under Grant FK 145838, by the J\'anos Bolyai Research Scholarship of the Hungarian Academy of Sciences, and by the Research Foundation - Flanders under Grant S005521N. \\
We are grateful to \emph{Andr\'as Gyimesi}, \emph{Sergey Ilyin}, and three anonymous reviewers for useful remarks.

\begingroup
\setstretch{1.1}
\setlength{\bibsep}{0pt plus 0.3ex}
\small{\bibliography{OR_survey_references}}
\endgroup

\end{document}